\documentclass{optica-article}

\journal{oe}

\newcommand{\Minus}{\raisebox{.3\height}{\resizebox{5.5pt}{2.5pt}{-}}} 

\articletype{Research Article}

\usepackage{makecell}
\usepackage{color,soul}
\usepackage{graphicx}
\usepackage{adjustbox}
\usepackage{caption}
\usepackage{siunitx,upgreek}

\begin{document}

\title{High-rate multiplexed entanglement source based on time-bin qubits for advanced quantum networks}

\author{Andrew Mueller\authormark{1, 2, *}, Samantha I. Davis\authormark{3}, Boris Korzh\authormark{2}, Raju Valivarthi\authormark{3}, Andrew D. Beyer\authormark{2}, Rahaf Youssef\authormark{3}, Neil Sinclair\authormark{3,4}, Cristián Peña\authormark{3,5}, Matthew D. Shaw\authormark{2}, and Maria Spiropulu\authormark{3}}

\address{\authormark{1}Applied Physics, California Institute of Technology, 1200 E California Blvd., Pasadena, CA, 91125, USA \\
\authormark{2}Jet Propulsion Laboratory, California Institute of Technology, 4800 Oak Grove Dr., Pasadena, CA, 91109, USA\\
\authormark{3}Division of Physics, Mathematics and Astronomy, California Institute of Technology, 1200 E California Blvd., Pasadena, CA 91125, USA \\
\authormark{4}John A. Paulson School of Engineering and Applied Sciences, Harvard University, 29 Oxford St., Cambridge, MA 02138, USA \\
\authormark{5}Fermi National Accelerator Laboratory, Batavia, Illinois 60510, USA}

\email{\authormark{*}amueller@caltech.edu} 

\begin{abstract}
Entanglement distribution based on time-bin qubits is an attractive option for emerging quantum networks. We demonstrate a 4.09 GHz repetition rate source of photon pairs entangled across early and late time bins separated by 80 ps. Simultaneous high rates and high visibilities are achieved through frequency multiplexing the spontaneous parametric down conversion output into 8 time-bin entangled channel pairs. We demonstrate entanglement visibilities as high as 99.4\%, total entanglement rates up to 3.55$\times 10^6$ coincidences/s, and predict a straightforward path towards achieving up to an order of magnitude improvement in rates without compromising visibility. Finally, we resolve the density matrices of the entangled states for each multiplexed channel and express distillable entanglement rates in ebit/s, thereby quantifying the tradeoff between visibility and coincidence rates that contributes to useful entanglement distribution. This source is a fundamental building block for high-rate entanglement-based quantum key distribution systems or advanced quantum networks. 
\end{abstract}

\section{Introduction}

Quantum computing represents an upcoming threat to public-key cryptography \cite{shor1997, nielsen2010quantum}. Quantum Key Distribution (QKD) is a method for overcoming this threat by sharing secret cryptographic keys between parties in a manner that is sufficiently secure against potential eavesdroppers and the decryption capabilities of quantum computers. Point-to-point QKD networks are a precursor to more advanced quantum networks which enable the transfer of quantum states for multiple applications including distributed quantum computing, sensing, or secure communication. We characterize any quantum network as ‘advanced’ if it enables protocols and capabilities that go beyond point-to-point QKD \cite{Wehner2018}. These include teleportation \cite{teleportation1993, Valivarthi2020}, entanglement swapping \cite{swapping1998}, memory-assisted networks \cite{Hermans2022}, and others. Entangled photons are a fundamental resource for such demonstrations, and entanglement distribution is therefore a key component of premier quantum network initiatives including the European Quantum Communication Infrastructure (EuroQCI) project, the Illinois Express Quantum Network (IEQNET), the Chinese Quantum Experiments at Space Scale (QUESS) initiative, the United Kingdom UKQNTel network, and the Washignton DC-QNet Research Consortium. Future quantum networks should enable high-fidelity and high-rate transfer of individual quantum states across multiple quantum nodes, mediated by distribution of entangled photons, quantum memories, and entanglement swapping measurements.



High-rate entanglement distribution enables high-rate entanglement-based QKD, as well as more general operations that characterize advanced quantum networks. Entanglement distribution and entanglement-based QKD have been demonstrated with impressive performance across a number of metrics. These include 40~kbps data rates in a QKD system deployed over 50~km of fiber\cite{Pelet2022} as well as multiple polarization entangled sources that leverage spectral multiplexing. These polarization sources include a demonstration of 181 kebits/s across 150 ITU channel pairs and a high-throughput source potentially capable of  gigabit rates with many added channels and detectors \cite{Alshowkan2022, Neumann2022Entanglement}. Multiple works have highlighted the need to  leverage high total brightness, spectral brightness, collection efficiency, and visibility from pair-generating non-linear crystals to realize practical high-rate entanglement distribution~\cite{Neumann2022Entanglement,atzeni2018integrated, sun2019compact,liu2021device,kaiser2014polarization, anwar2021entangled,neumann2021model}. 

A time-bin entangled photon source has certain advantages over a polarization-based system~\cite{marcikic2002time}. Time-bin entanglement can be measured with no moving hardware and does not require precise polarization tracking to maximize visibility~\cite{Dong2018PolarizationControll, Fitzke2022TimeBinVsPol}. Also, with suitable equipment, robust time-bin modulation is possible over free space links with turbulence~\cite{Jin2019}. Therefore, the possibility of simplified fiber-to-free-space interconnects and larger quantum networks based on a shared time-bin protocol motivates development of improved time-bin sources. Furthermore, time-bin encoding is suited for single-polarization light-matter interfaces \cite{simon2010quantum}.

\begin{figure}
    \centering
    \includegraphics[width=1\linewidth]{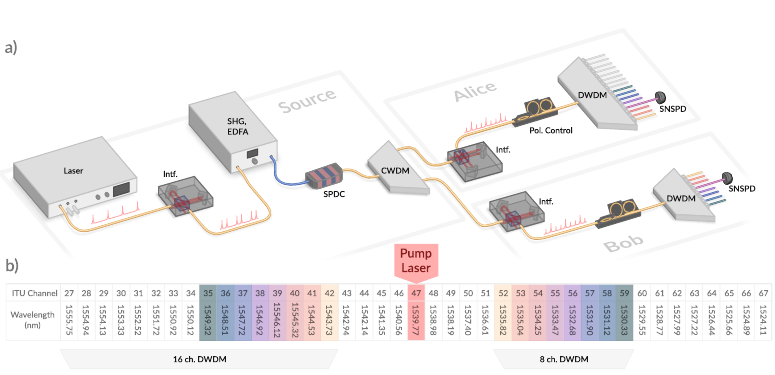}
    \caption{a) Pulses from a 1539.47 nm mode locked laser (Pritel UOC) are split into two by an 80-ps delay-line interferometer before up-conversion and amplification in a second harmonic generation + erbium doped fiber amplifier (SHG + EDFA) module (Pritel). A short PM fiber from the SHG module connects to a nonlinear crystal generating photon pairs by spontaneous parametric down-conversion (SPDC). The coarse wavelength division multiplexing (CWDM) module separates the photon pair spectrum into eight 13~nm-wide bands around 1530 and 1550 nm, for the signal and idler photon, respectively. The signal and idler are directed to the Bob and Alice stations, respectively. The readout interferometers introduce the same time delay as the source interferometer. Polarization controllers are used to maximize the coincidence rates, as the detection efficiencies of each SNSPD is polarization sensitive ($\pm10\%$). Entanglement visibility is unaffected by readout polarization. The polarization controllers could be removed if future systems adopt polarization insensitive SNSPDs \cite{Mukhtarova18}. 100 GHz spacing dense wavelength division multiplexer (DWDM) modules are used to direct each frequency channel into a distinct fiber. Two superconducting nanowire single photon detectors (SNSPDs) are used to measure a specific frequency multiplexed channel pair. Measurements for different multiplexed channels are performed in succession to resolve full system performance. b) ITU channels used in the experiment. Pairs of channels highlighted with the same color obey the phase and pump-energy matching condition for SPDC. To asses the full 16 channels (27-42) of Alice's DWDM multiplexer, Bob's 8-channel DWDM is replaced with a narrowband filter with tunable resonance frequency (not shown in figure).}
    \label{fig:system}
\end{figure}

We direct 4.09 GHz mode locked laser light into a nonlinear crystal via 80-ps delay-line interferometers (12.5 GHz free-spectral range) to realize a high-rate entanglement source. The ability to resolve time-bin qubits into 80~ps wide bins is enabled by newly developed low-jitter differential superconducting nanowire single-photon detectors (SNSPDs)~\cite{Colangelo2023}. Wavelength multiplexing is used to realize multiple high visibility channel pairings which together sum to a high coincidence rate. Each of the pairings can be considered an independent carrier of photonic entanglement \cite{Djeylan2016, Wengerowsky2018} and therefore the system as a whole is applicable to flex-grid architectures through the use of wavelength selective switching \cite{Appas2021, Alshowkan22Switching}. However, we focus on maximizing the rate between two receiving stations, Alice and Bob (Fig. \ref{fig:system}a). Each station is equipped with a DWDM that separates the frequency multiplexed channel into multiple fibers for detection. The SNSPDs are used with a real-time pulse pileup and time-walk correction technique~\cite{Mueller2023} to keep jitter low even at high count rates.

We quantify per-channel brightness and visibility as a function of pump power, as well as collection efficiencies, coincidence rates across 8 channel pairs. Performance of a 16-channel pair configuration is discussed in Supplementary note 4. We show that the 8-channel system achieves visibilities that average to 99.3\% at low mean photon number $\mu_{L} = 5.6{\times} 10^{\Minus5}\,\pm\,9{\times} 10^{\Minus6}$. At a higher power ($\mu_{H} = 5.0{\times} 10^{\Minus3}\,\pm\,3{\times} 10^{\Minus4}$), we demonstrate a total coincidence rate of 3.55 MHz with visibilities that average to 96.6\%. Through quantum state tomography we bound the distillable entanglement rate of the system to between 69\% and 91\% of the $\mu_{H}$ coincidence rate (2.46 - 3.25 Mebits/s).

Quantifying a source's spectral mode purity is important for gauging its utility in advanced quantum networks that rely on interferometric measurements like two-photon interference which enables Bell-state measurements (BSM) \cite{Valivarthi2020}. 
With Schmidt decomposition we quantify the modal purity of single DWDM channel pairs and derive the inverse Schmidt number which serves as an estimate for two-photon interference visibility between two such sources. Ultimately, we demonstrate that an entanglement generation source of this design makes for a robust and powerful building block for future high-rate quantum networks. 

\section{System}
Figure~\ref{fig:system} shows the experimental setup. Pulses from the 4.09 GHz mode-locked laser, with a center wavelength at 1539.47 nm, are sent through an 80 ps delay-line interferometer (Optoplex DPSK Phase Demodulator). All interferometers used are the same type; they have insertion loss of $1.37~\pm~0.29~\mathrm{dB}$, are polarization independent, and have extinction ratios greater than 18~dB. The source interferometer produces two pulses each clock cycle used to encode early/late basis states ($|e\rangle,|l\rangle$), which are subsequently up-converted by a second harmonic generation (SHG) module (Pritel) and down-converted into entangled photon pairs by a type-0 spontaneous parametric down conversion (SPDC) crystal (Covesion) \cite{Marcikic2002}. The SPDC module uses a 1~cm long waveguide-coupled MgO-doped lithium niobate crystal with an 18.3 $\unit{\micro\meter}$ polling period. The up-converted pulses at 769 nm have a FWHM bandwidth of 243 GHz (0.48 nm), which along with the phase matching condition of the SPDC waveguide, defines a wide joint spectral intensity (JSI) function \cite{kim2005measurement}. 

The photon pairs are separated by a coarse wavelength division multiplexer (CWDM) which serves to split the SPDC spectrum into two wide-bandwidth halves. For a system using more than 16 DWDM channels at Alice and Bob, the CWDM would be replaced with a splitter that efficiently sends the full SPDC spectrum shorter than 1540~nm to Bob, and the spectrum longer than 1540 nm to Alice. A dichroic splitter with a sharp transition at 1540~nm would also enable the use of DWDM channels 43-46 and 48-51. The pairs are of the form $|\psi\rangle=\frac{1}{\sqrt{2}}\left(|e\rangle_{s}|e\rangle_{i}+e^{i \phi}|l\rangle_{s}|l\rangle_{i}\right)$. Entangled idler and signal photons are sent to the receiving stations labeled Alice and Bob, respectively. One readout interferometer at each station projects all spectral bands into a composite time-phase basis. From here, dense wavelength division multiplexers (DWDM) divide up the energy-time entangled photon pairs into spectral channels. 

DWDM outputs are sent to differential niobium nitride (NbN) single pixel SNSPDs~\cite{Colangelo2023} with 22 × 15~µm active areas formed by meanders of 100-nm-wide and 5-nm-thick niobium nitride (NbN) nanowires on a 500~nm pitch. These measure the arrival time of photons with respect to a clock signal derived from the mode locked laser. Use of the high system repetition rate and compact 80~ps delay interferometers is only possible due to the high timing resolution of these detectors. Low jitter performance is achieved by incorporating impedance matching tapers for efficient RF coupling, resulting in higher slew rate pulses, and by enabling RF pulse readout from both ends of the nanowire. The dual-ended readout allows for the cancellation of jitter caused by the variable location of photon arrival along the meander when the differential signals are recombined with a balun. SNSPDs of this type reach system jitters down to 13.0~ps FWHM, and 47.6~ps FW(1/100)M \cite{Colangelo2023}. We use two SNSPDs for this demonstration with efficiencies at 1550~nm of 66\% and 74\%. They exhibit 3~dB maximum count rates of 15.1 and 16.0~MHz. A full 8-channel implementation of this system would require 16 detectors operating in parallel at both Alice and Bob. To read out both outputs of both interferometers, 4 detectors per channel are required, resulting in 32 detectors total. 

A novel time-walk or pulse-pileup correction technique is used to extract accurate measurements of SNSPD pulses that arrive between 23 and 200~ns after a previous detection on the same RF channel. Without special handling of these events, timing jitter will suffer due to RF pulse amplitude variations and pileup effects. As detailed in Supplementary note 3, the correction method works by subtracting off predicable timing distortions based on the inter-arrival time that precedes them~\cite{Mueller2023, Craiciu23}. An in-situ calibration process is used to build a lookup table that relates corrections and inter-arrival time. At the highest achievable pump power, this correction method leads to 320\% higher coincidence rates compared to a data filtering method that rejects all distorted events arriving within $\simeq$200~ns of a previous pulse. 

\section{Results}

\begin{figure}
    \centering
    \includegraphics[width=0.65\linewidth]{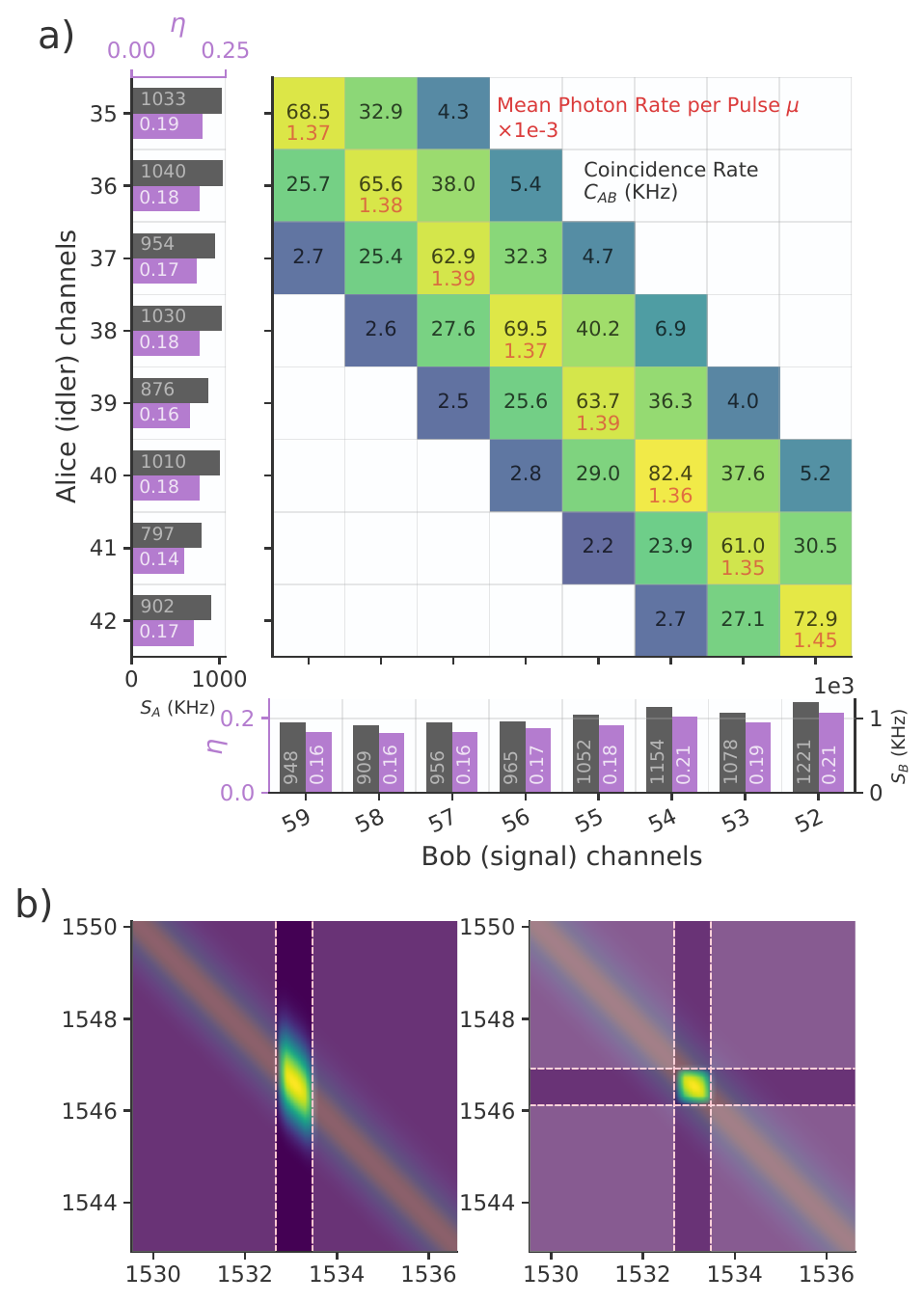}
    \caption{a) The singles rates at Alice $S_A$ and Bob $S_B$ (grey bars), path coupling efficiencies (purple bars), and coincidence rates $C_{AB}$ (black text in colored boxes) for different DWDM channel pairings. All measurements are recorded at a SHG pump power of 14.6~mW, for which $\mu$ of the channel pairs along the main diagonal are shown in red text. The joint spectral intensity envelop spans several 100 GHz channels. As detailed in Supplementary note 2, the coincidence rates (kHz, black) are scaled to represent two branches of the total wavefunction, so that they are consistent with the coupling efficiencies $\eta$ (purple bars) and the singles rates (grey bars). There are four branches for each pairing of the four interferometer output ports. In practice, one output each of Alice and Bob's interferometers is measured, thereby capturing one branch. See Supplementary note 6 for details of the fitting method used to solve for the coupling efficiencies $\eta$. b) Visualization used to motivate the geometric factor $\delta$. The portion of the JSI that is captured by one 100~GHz filter (singles) is on the left, and that captured by two filters (coincidences) is on the right.  Assuming 100\% coupling efficiency inside the filter passbands, the ratio of captured coincidences on the right over singles on the left is $\delta$. For context, the portion of the JSI that is filtered away is faintly visible.}
    \label{fig:figure_2nd_1}
\end{figure}

By pairing up particular 100~GHz DWDM channels and recording coincidence rates, a discretized form of the JSI of our pair source is measured (Fig.~\ref{fig:figure_2nd_1}a). Due to the wide pump bandwidth, the spectrum of signal photons spans several ITU channels for a given idler photon wavelength. Pairs along the main diagonal are optimized for maximum coincidence rates by tuning the pump laser frequency, and are therefore used for all remaining measurements. In Fig. \ref{fig:system}b, these pairs are highlighted with matching colors. Coupling or heralding efficiencies $\eta$ shown in Fig.~\ref{fig:figure_2nd_1}a are derived from a JSI fitting analysis (see Supplementary note 6) and include all losses between the generation of entangled pairs in the SPDC and final photo-detection.

The mean pair rate is commonly derived as $\mu = S_A S_B / (R C_{AB})$, where $S_A, S_B$ are the detector count rates at Alice and Bob, $C_{AB}$ is the coincidence rate, and $R$ is the system repetition rate. This is appropriate when losses on the signal and idler arm do not significantly depend on wavelength. But the formula gives misleading $\mu$ values when DWDM channel filtering is narrowband relative to the width of the JSI. In this experiment only a fraction of idler (signal) photons that exit one DWDM channel will be detected with their corresponding signal (idler) photon, even for ideal DWDM channels with 100\% transmission within their passbands. This property is illustrated in Fig.~\ref{fig:figure_2nd_1}d. This implies a geometric limit on the ratio of coincidence to singles rates in this narrowband multiplexing regime. We account for this in calculations of $\mu$ by adding a geometric compensation factor $\delta$ to the commonly used equation:


\begin{align}
\mu =  \frac{\delta S_A S_B}{R C_{AB}}.
\end{align}

This gives a definition of $\mu$ for the JSI region where signal and idler filters overlap according to energy conservation, and the probability of transmitting entangled pairs to both Alice and Bob is non-negligible. It is valid in the low $\mu$ regime where generation of higher order photon number states from the SPDC are rare. For filter pairings along the main diagonal in Fig.~\ref{fig:figure_2nd_1}a, values for $\delta$ are fairly consistent and average to $\delta = 0.393 \pm 0.012$. Related experiments have derived $\mu$ from a Coincidence to Accidental Ratio (CAR) measurement. Here, a typical CAR measurement is misleading given the narrowband filter regime, and it does not align with our unique definition of $\mu$. Further motivation and derivation of $\delta$ is included in Supplementary note 7.



\begin{figure}
    \centering
    \includegraphics[width=0.65\linewidth]{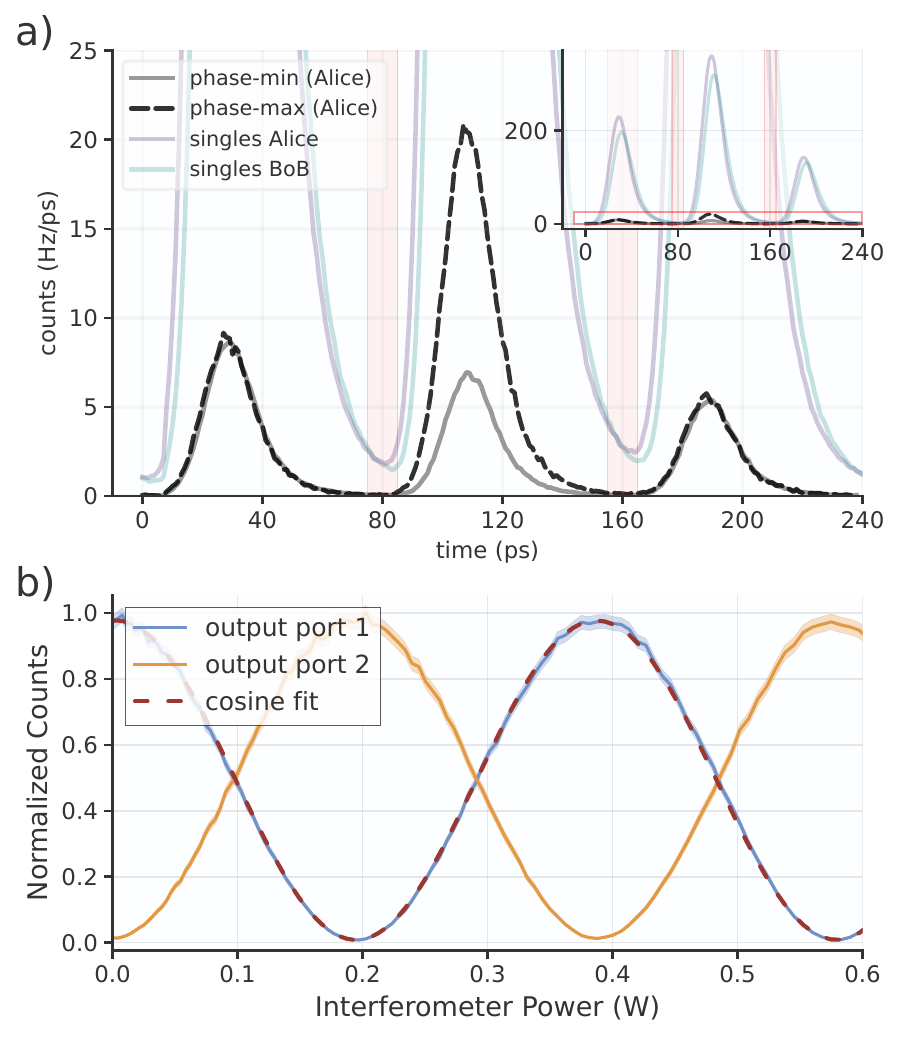}
    \caption{a) Histogram of photon arrival events with respect to the 4.09 GHz clock. Dashed black and grey lines show the response functions for coincidence events. Events within 10~ps guard regions centered at 80 and 160~ps (shaded red) are discarded for analysis of coincidences between individual bins. This is done to maximize visibility in the presence of some minor overlap of the pulses (see Supplementary note 5 for discussion). The coincidence histograms include pairings from any combination of early, middle, and late time bins. Therefore, the height of the center peak in the phase-min state is not near zero, as non-phase-varying terms contribute. b) Coincidence rate interference fringes for the center time bin in isolation. Based on the good agreement between the fringe data and a cosine fit, we make subsequent tomographic measurements assuming that phase is linear with the electrical power applied to the interferometer phase shifter.}
    \label{fig:figure_2nd_2}
\end{figure}

In the following, rigorous tests of entanglement are primarily done with the 8 ITU 100 GHz channel pairings: Ch. 35-42 at Alice and Ch. 52-59 at Bob. However, as shown in Supplementary note 4, source brightness measurements were conducted on a partially realized 16-channel configuration which makes use of all 16 channels available on Alice's DWDM. 

Signals from the SNSPDs are directed to a free-running time tagger (Swabian) and processed with custom software. The resulting histograms, referenced from a shared clock (Fig \ref{fig:figure_2nd_2}a), depict three peaks, which are caused by the sequential delays of the source and readout interferometers. Some intensity imbalance between long and short paths is present in these interferometers, which explains the asymmetry between early and late peaks in Fig. \ref{fig:figure_2nd_2}a. Such imbalances are present in both the source and readout interferometers to varying degrees. The interferometer used for the source exhibits an early/late intensity balance ratio of 1.13. Alice and Bob's interferometers exhibit early/late imbalances of 1.24 and 1.15 respectively. These induce imperfect overlap of certain time-bin modes of differing amplitudes. This mismatch lowers interference visibilities, as detailed in Supplementary note 10.

The coincidence rate across Alice and Bob's middle bins varies sinusoidally with respect to the combined phase relationship of the source and readout interferometers (see Supplementary note 1) \cite{Inagaki2013, Marcikic2002}. In Fig. \ref{fig:figure_2nd_2}a  the coincidences shown are for any combination of early, middle, or late bins. For tomography and visibility measurements, coincidence detections across specific bin pairings are considered.  

\begin{figure}
    \centering
    \includegraphics[width=0.8\linewidth]{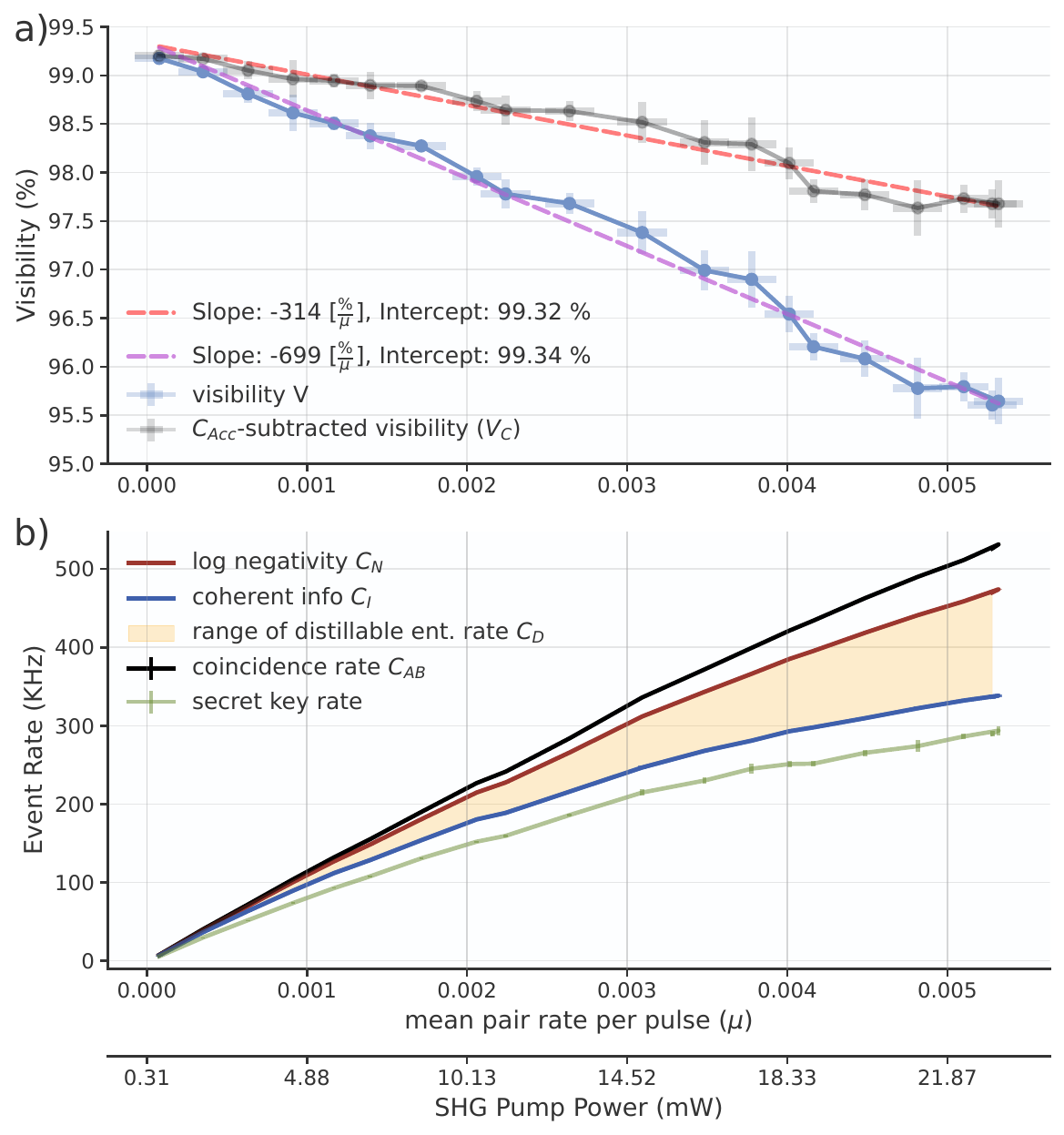}
    \caption{a) Visibility versus pump power. Error bars are calculated by taking multiple measurements of the center bin coincidence rate over some integration time. These measurements span small ranges of interferometer phase, as the extremum-finding algorithm jitters the interferometer voltage. $V_C$ (grey data, red line) is a construction that models how visibility would be affected if accidental coincidences from mutually incompatible spectral modes could be mitigated in future systems. b) Bounded distillable entanglement rate versus pump power. Multiple such measurements are made for all the tomographic measurements. These are used to calculate standard deviations for visibility, log negativity, and coherent information. Error bars for the log negativity and coherent information are smaller than the line width shown. Rates shown assume readout of all 4 available interferometer ports, based on data measured using one port each at Alice and Bob.}
    \label{fig:shg_scan}
\end{figure}

Due to the small size ($3~\times~3~\mathrm{cm}$) and temperature insensitivity of the interferometers, minimal temporal phase drift is observed. Without active temperature control or phase feedback, we observe minimized coincidence rates of the center time bin stay within 6\% of their original values after 50 minutes. Nevertheless, software is used to lock the voltage-controlled phase at a minimum or maximum with a simple hill-climbing algorithm. This varies the phase by small amounts over several minutes to search for or maintain an extremum. This is simpler to implement than the techniques needed to stabilize interferometers of longer path length difference, including the use of precise temperature control \cite{Valivarthi2020} or co-propagating stabilization lasers \cite{Toliver15}.

Channels 35 and 59 are chosen for an analysis of entanglement visibility and rates versus pump power. Visibility with respect to pump power or mean entangled pair rate is shown in Fig.~\ref{fig:shg_scan}a. We define the entanglement visibility as $V = 100\%*(C_{max} - C_{min})/(C_{max} + C_{min})$ where $C_{min}$ and $C_{max}$ are the minimum and maximum coincidence rates in the middle bin for varied phase. As this coincidence rate depends on the total phase across the source and readout interferometers, only Bob's interferometer is actively controlled to scan the full state space. 

The raw visibility versus $\mu$ is shown in blue in Fig.~\ref{fig:shg_scan}a. Relative to similar measurements \cite{Kim2022}, this drops quickly with increasing $\mu$, and one reason is the presence of accidental coincidences across mutually incompatible spectral modes. The presence of these unwanted coincidences is a consequence of the narrowband filtering regime, and depends on factors included the singles rates $S_A$ and $S_B$, and the geometric compensation factor $\delta$ (see Supplementary note 9 for derivation). We model this type of accidental coincidence rate $C_{Acc}$ versus $\mu$, and subtract it off from coincidence measurements to produce the grey data in Fig.~\ref{fig:shg_scan}a. This simulated visibility's more gradual drop with increasing $\mu$ highlights the detrimental effect of our high single-to-coincidence rates $S_A/C_{AB}$, $S_B/C_{AB}$. As detailed in the discussion section below, this motivates special source engineering techniques for future systems. Supplementary note 12 infers the auto-correlation function $g^2(0)$ vs $\mu$ for this source in the small $\mu$ limit.

\begin{figure}
    \centering
    \includegraphics[width=0.8\linewidth]{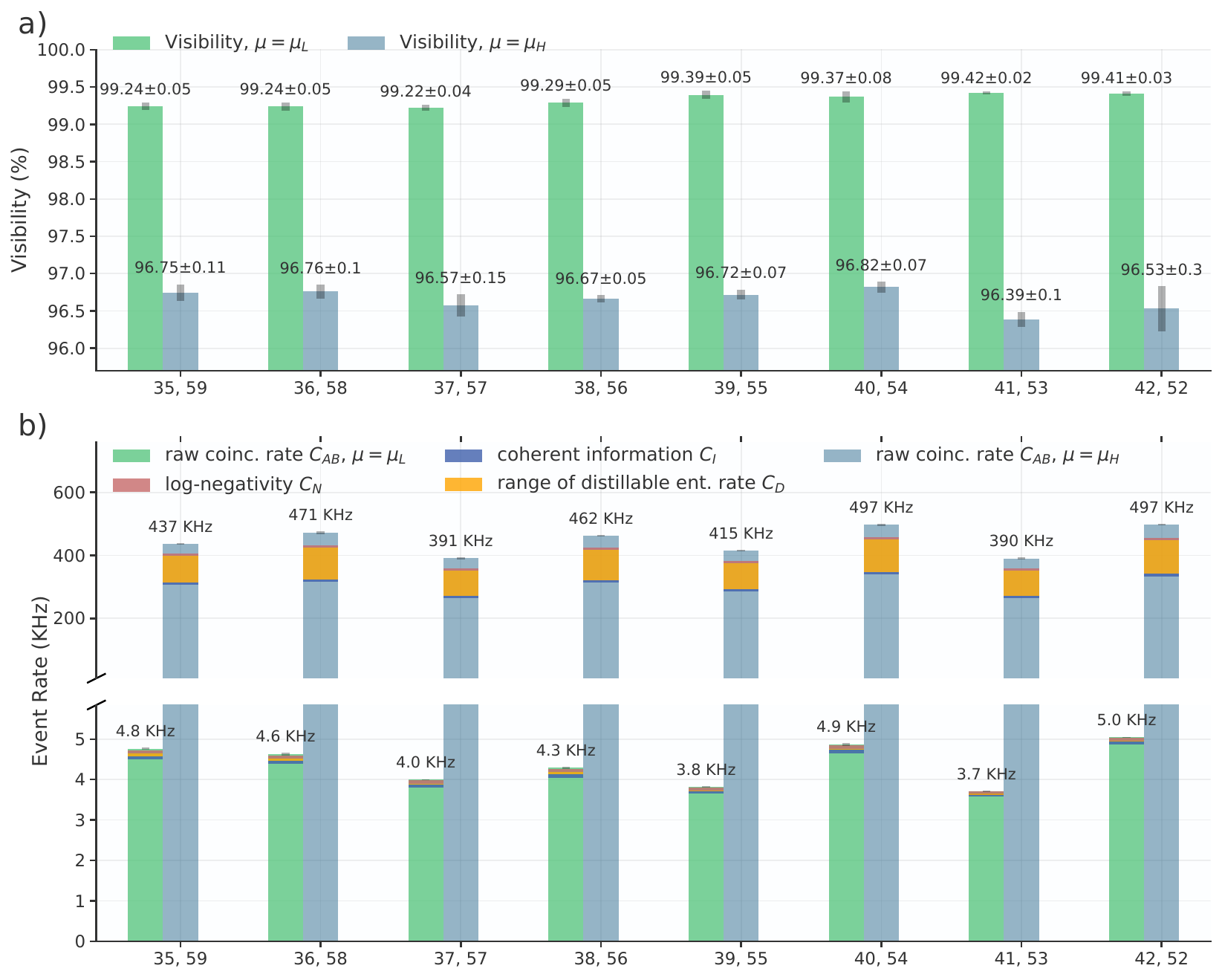}
    \caption{a) Visibility for the main 8 channel pairs, measured at a high (22.9~mW) and a low (0.21~mW) SHG pump power setting. Each power setting results in similar $\mu$ for all channels: $\mu_L = 5.6{\times}10^{-5} \pm 9{\times}10^{-6}$ and $\mu_H = 5.0{\times}10^{-3} \pm 3{\times}10^{-4}$.  b) Rate metrics for the 8 channel pairs at the same high and low power settings. The range of possible values for distillable entanglement rate is spanned by the yellow regions, bounded above by log-negativity and below by coherent information. Rates shown assume readout of all 4 available interferometer ports, based on data measured using one port each at Alice and Bob}
    \label{fig:channel_data}
\end{figure}

We quantify the rate of useful entanglement by supplying bounds for the distillable entanglement rate $C_D$. Measured in ebits/s, $C_D$ is the maximal asymptotic rate of Bell-pair production per coincidence using only local operations and classical communications \cite{Alshowkan2022, Bennett1996}. It is bounded above by log-negativity $C_N = C_{AB} E_N$ and below by coherent information $C_I = C_{AB} E_I$ \cite{Alshowkan2022}. For each pump power setting in Fig.~\ref{fig:shg_scan}, a series of tomographic measurements is performed and density matrices are calculated. The values of $E_I$ and $E_N$ are calculated from the density matrices as detailed in Supplementary note 8. 

Figure \ref{fig:channel_data} shows visibilities, raw coincidence rates, and bounded distillable entanglement rates for two pump powers and all 8 channel pairings. The highest pump power is currently limited by our EDFA-amplified SHG module. The pump power in principle could be increased until the SNSPD efficiency drops due to saturation, and the net coincidence rate plateaus. Without the time-walk correction, high-rate jitter becomes an issue well before the gradual drop of SNSPD efficiency. At the $\mu_H$ (22.9~mW) power, the singles rates $S_A, S_B$ average to 3.84 MHz, for which SNSPD efficiencies are about 78\% of nominal. 


Using the data in Fig.~\ref{fig:figure_2nd_1}a, we model the JSI of our pair source as a product of pump envelope and phase matching condition functions

$$|f(\omega_s, \omega_i)|^2 = |\psi_{\mathrm{ph}}\left(\omega_s, \omega_i\right)|^2 *|\psi_p\left(\omega_s, \omega_i\right)|^2,$$
which depends on the wavelength (769.78 nm) and bandwidth (243 GHz FWHM) of up-converted light out of the SHG, measured with a spectrum analyzer. The path efficiencies from SPDC to detectors are also fitted based on integrations over the JSI that model the DWDM transmission passbands (see Supplementary note 6). 

We calculate the Schmidt decomposition of the pair source JSI, taking into account the DWDM filters at Alice and Bob, and derive an average inverse Schmidt number $1/K$ of $0.87$. This value quantifies the spectral purity of the entangled photon source, and is theoretically equivalent to the visibility of a two-source HOM (Hong-Ou-Mandel) interferogram~\cite{mandel1995optical}. If 50 GHz ITU channels are used instead, the resulting filtered JSI better approximates a single mode, and the model predicts $1/K = 0.96$. The increase of $1/K$ with tighter filtering, as predicted by our model, is supported by simpler analytical expressions \cite{ZielnickiKwiat2018SPDCmodel}. Relative to the 100~GHz channels used here, we expect the use of 50~GHz channels to decrease singles rate by roughly a factor of two, and coincidence rate by a factor of four for equivalent pump power. Our definition of $\mu$ would require a larger geometric factor $\delta$, and visibility would therefore scale less favorably with $\mu$ due to the greater impact of accidental coincidences. Ultimately, due to this narrowband filtering regime and geometric filter considerations, there are trade-offs to using narrower filters which may be acceptable for some applications, but not others.

\section{Discussion}


\begin{table}
\small
\caption{Architecture and Rates Comparison} \label{table:rates_comparison}
\begin{adjustbox}{center}
 
\begin{tabular}{ccccccc}
  
  \cline{1-7}
  Refs. &
  DOF &
  \makecell[c]{ch. pairs tested \\ \textcolor{blue}{(type)}}  &
  \makecell[c]{coinc. rate (Hz) \\ total \textcolor{blue}{(avg. per ch.)}} &
  \makecell[c]{$C_D$, total (ebits/s) \\ \textcolor{blue}{(avg. per ch.)}} & 
  \makecell[c]{SKR, total (Hz) \\ \textcolor{blue}{(avg. per ch.)}} & \makecell[c]{brightness \\ $(\mathrm{Hz}/\mathrm{mW}/\mathrm{nm})$} \\ \hline
  
  Neumann (2022)\cite{Neumann2022Entanglement} & pol. & 7, 1 \textcolor{blue}{(200,100 GHz)} & $\sim$1.2e5 \textcolor{blue}{(1.6e4)} & - & $\sim$1.6e5 \textcolor{blue}{(2.2e4)} & $\sim$2.7e5 \\

  Alshowkan (2022)\cite{Alshowkan2022} & pol. & 150 \textcolor{blue}{(25 GHz)} & >2.01e5  \textcolor{blue}{(>1.3e3)} & \makecell[c]{1.81e5 - 2.01e5 \\ \textcolor{blue}{(1210 - 1340)}} & - & >5.3e3 \\
  
  Kim (2022)\cite{Kim2022} & time & 3 \textcolor{blue}{(100 GHz)} & \makecell[c]{54.6 \textcolor{blue}{(18.2)}} & - & 11.47 \textcolor{blue}{(3.82)} & - \\

  Wengerowsky (2018) \cite{wengerowsky2018entanglement} & pol. & 6 \textcolor{blue}{(100 GHz)} & 311 \textcolor{blue}{(52.0)} & -  & $\sim$1.1e2 \textcolor{blue}{(20)} & - \\

  Aktras (2016) \cite{Aktras2016} & time & 8 \textcolor{blue}{(100 GHz)} & 3.8e4 \textcolor{blue}{(4.5e3)} & - & - & - \\
  
  this work & time & 8 \textcolor{blue}{(100 GHz)} & \makecell[c]{3.55e6 \textcolor{blue}{(4.43e5)}} & \makecell[c]{2.46e6 - 3.25e6 \\ \textcolor{blue}{(3.08e5 - 4.06e5)}} & 1.96e6 \textcolor{blue}{(2.40e5)} & 1.3e5 \\ \hline
\end{tabular}
\end{adjustbox}
\caption*{Table \ref{table:rates_comparison}. Architectures and rate metrics for other wavelength multiplexed entanglement distribution sources. Neumann (2022) \cite{Neumann2022Entanglement} uses 7 channel pairs of 200 GHz width and one pair of 100 GHz width. Per-channel rate is estimated for the 200 GHz filters. Brightness is based on measured coincidence rate. The coincidence rate for Aktras (2016) \cite{Aktras2016} is at a visibility of 83\%. Tilde ($\sim$) indicates computed estimates based on available data. While some sources also included rate metrics with added lengths of optical fiber between source and readout, those are not presented here. }
\end{table}


We have demonstrated that a time-bin entanglement source based on a mode-locked laser, spectral multiplexing and low-jitter detectors produces high entangled photon rates suitable for QKD or advanced quantum networks. The distillable entanglement rate, achievable secret key rate, and visiblilities of this source are highly competitive relative to other multiplexed entanglement distribution systems (Table \ref{table:rates_comparison}). Still, there is potential to increase rates beyond those measured here with some straightforward changes to the setup. First, a higher power EDFA-amplified SHG module or tapered amplifier may be used. With this, we predict a single channel pair could sustain rates up to those specified in the first column of Table \ref{table:max_rates}. These metrics all depend on both entanglement quality and coincidence rate $C_{AB}$. Due to the trade-off between $C_{AB}$ and entanglement quality or visibility, they all reach maximum values for particular pump powers extrapolated in Supplementary note 11. Our measurements of 8 channel and 16 channel configurations imply the approximately multiplicative scalings in columns 2 and 3 of Table \ref{table:max_rates}, as coincidence rates of these channels pairs are all withing 27\% of each other.  From measurements of the SPDC spectrum, it is also possible to extrapolate rates to a 60-channel 100 GHz DWDM configuration that includes channels spanning the L, C, and S ITU bands. This configuration could sustain 34.9 MHz total coincidence rate, and a distillable entanglement rate between 27.7 ($C_N$) and 15.9 Mebits/s ($C_I$).  These rates are impressive considering they are achievable with existing SNSPDs and other technology. A measurement of the full SPDC spectrum and extrapolation details are found in Supplementary note 11.

\begin{table}
\small
\captionof{table}{Extrapolated rates (MHz)} \label{table:maximum_rates} 
\begin{tabular}{ccccc}
 \cline{1-3}
 \hline
 \makecell[l]{rate metric ($\mu$ at max)}   &  1 Channel & 8 Channels & 16 Channels & 60 Channels \label{table:max_rates}\\
 \hline
 \makecell[l]{coincidence rate, $C_{AB}$ (0.014)} & 0.755 & 5.41    &  11.6  &  34.9     \\
 \makecell[l]{log negativity, $C_N$ (0.010)}& 0.600 & 4.30    &  9.19  &  27.7   \\
 \makecell[l]{coherent info., $C_I$ (0.006)}& 0.345 & 2.47 &  5.28  &  15.9 \\
 \makecell[l]{secret key rate, $SKR$ (0.007)}& 0.309 & 2.21 &  4.73  &  14.3 \\
 \hline
\end{tabular}
\caption*{Table \ref{table:maximum_rates}. Per-channel predicted maximum values for the 4 rate metrics are shown in the `1 Channel' column. Depending on the metric, the maxima are achieved for different pump powers $\mu$ (See supplemental note 11). The $\mu$ value that maximizes each metric is shown in parenthesis on the left.}
\end{table}

The ratio of singles rates $S_A, S_B$ to coincidence rates $C_{AB}$ are high in this system due to the relatively wide-band JSI and narrow filters. Each DWDM channel at Alice picks up a large fraction of photons that can't be matched with pairs passing though the corresponding channel passband at Bob, a feature quantified by the $\delta$ factor. The high singles rates lead to accidental coincidences from mutually incompatible spectral modes that lower visibility and load the detectors with useless counts. However, there is potential to mitigate these extra counts by embedding the nonlinear crystal undergoing SPDC in a cavity that enhances emission at the center frequencies of multiple DWDM channels~\cite{Pomarico2009, Brydges2023, slattery2019background}. Also, there are other approaches to achieving such intensity islands that require dispersion engineering \cite{morrison2022frequency, xin2022spectrally}. With such periodically enhanced emission, the resulting JSI would exhibit a series of intensity islands lying along the energy-matching anti-diagonal, easily separable with DWDMs at Alice and Bob. The photon flux for each each channel would originate primarily from these islands covered by both signal and idler DWDM passbands, resulting in a higher ratio of coincidences to singles. The probability of accidental coincidences $C_{Acc}$ would be lower, and therefore bring the decrease of visibility with $\mu$ more in line with the modeled $V_C$ data in Fig.~\ref{fig:shg_scan}. We intend for the $V_C$ construction to represent how visibility would degrade primarily due to multi-pair effects, assuming accidental coincidences from incompatible spectral modes could be mitigated. The more gradual decrease in visibility with $\mu$ would enable substantially higher maximum rate metrics than those in Table \ref{table:maximum_rates}. 

This source is a fundamental building block for future space-to-ground and ground-based quantum networks. It leverages the strengths of the latest SNSPD developments -- namely simultaneous high count rates, low jitter and high efficiency -- and in doing so adopts interferometers and DWDM systems that are compact, stable and accessible. By elevating the system clock rate to 4.09 GHz and shrinking the time bin size to 80~ps, we have demonstrated a new state of the art in quantum communication that enables adoption of mature and extensively developed technologies from classical optical networks. Also, the spectral multiplexing methods used here are potentially compatible with those demonstrated in broadband quantum memories~\cite{Sinclair2014} and optical quantum computing~\cite{lukens2017frequency}.



\section*{Acknowledgements}
Part of the research was carried out at the Jet Propulsion Laboratory, California Institute of Technology, under a contract with the National Aeronautics and Space Administration (NASA) (No. 80 NM0018D0004). Support for this work was provided in part by the Defense Advanced Research Projects Agency (DARPA) Defense Sciences Office (DSO) Invisible Headlights program, NASA SCaN, Alliance for Quantum Technologies’ (AQT) Intelligent Quantum Networks and Technologies (INQNET) program, and the Caltech/JPL PDRDF program. A. M. is supported in part by the Brinson Foundation and the Fermilab Quantum Institute. M.S. is in part supported by the Department of Energy under Grant Nos. SC0019219 and SC002376. We are grateful to Si
Xie (Caltech/Fermilab) and Cristi{\' a}n Pe{\~ n}a (Fermilab) for supporting this work in terms of sharing equipment and facilities. The authors acknowledge Prathwiraj Umesh for assistance in reviewing the manuscript. 

\section*{Disclosures}
The authors declare no conflicts of interest.

\section{Data availability}
Data underlying the results presented in this paper are available in \cite{mueller2023code, mueller2023data}

\bibliography{references}

\end{document}


\maketitle

\section{Phase basis readout}

This section uses numbers in kets to signify time delays, such that the notation from the main text transforms as $|e\rangle, |l\rangle \longrightarrow |0\rangle, |1\rangle$. Following the creation of the bell pair $\frac{1}{\sqrt{2}}(|00\rangle + e^{i \phi}|11\rangle)$ with the source interferometer, the readout interferometers at Alice and Bob transform each member of the entangled pair according to the operation \cite{Marcikic2002}:

$$|k\rangle \rightarrow \frac{1}{2}\left(|k\rangle_{(A/B)+}+e^{i \phi_{s / i} \mid}|k+1\rangle_{(A/B)\Plus}+i|k\rangle_{(A/B)\Minus}-i e^{i \phi_{s / i}}|k+1\rangle_{(A/B)\Minus}\right)$$
Where $(A/B)\Plus$ and $(A/B)\Minus$ denote the output ports of Alice ($A$) or Bob's ($B$) interferometer. The full state is a 28-term expression made of four so-called 'branches' indexed by the four combinations of interferometer output ports: $A\Plus B\Plus, A\Plus B\Minus, A\Minus B\Plus$ and $A\Minus B\Minus$. Each branch has a term in the following form, with amplitude dependent on the phase relationship between the interferometers:

$$p\left(e^{i \phi}+qe^{i\left(\phi_s+\phi_i\right)}\right)|2\rangle_{Au}|2\rangle_{Bv} $$

Where $p, q, u, v \in \{\{\Plus1,\Plus1, \Plus, \Plus\}, \{i,\Minus1, \Plus, \Minus\}, \{i,\Minus1, \Minus, \Plus\}, \{\Minus1,\Plus1, \Minus, \Minus\}\}$ for the four terms. These terms define the probability amplitude of the quantum state in the so-called phase basis. The modulous squared of these terms gives the phase-dependent probability of coincidences across the center time bins, as measured at interferometer outputs of Alice and Bob. 

All the phase terms can be grouped into one variable $\theta = \phi_s + \phi_i - \phi$, and the four $|2\rangle|2\rangle$ terms can be re-expressed in terms of the the cosine function \cite{Marcikic2002, Kim2022}:

\begin{align}
P_{A\Plus B\Plus} &= |\langle 2|2\rangle|^2_{A\Plus B\Plus} = 2(1 + v \cos(\theta))  \label{cosines}\\
P_{A\Plus B\Minus} &= |\langle 2|2\rangle|^2_{A\Plus B\Minus} = 2(1 - v \cos(\theta)) \notag \\
P_{A\Minus B\Plus} &= |\langle 2|2\rangle|^2_{A\Minus B\Plus} = 2(1 - v \cos(\theta)) \notag \\
P_{A\Minus B\Minus} &= |\langle 2|2\rangle|^2_{A\Minus B\Minus} = 2(1 + v \cos(\theta)) \notag \\ \notag
\end{align}
where $v$ was added to denote the visibility of the phase basis. Scanning the phase of the system and measuring coincidence rate across the center bins produces sinusoidal fringes as shown in \ref{fig:fringes}.

\section{Coincidence Rate \& Interferometer Output Ports}

As each readout interferometer has two output ports, the full output state observed at Alice and Bob cannot be fully measured with two SNSPDs. We label the output ports of Alice (A) and Bob's (B) interferometers with plus ($\Plus$) and minus ($\Minus$). The plus ports are used for most measurements. By measuring the relative loss between the plus and minus ports, all singles rates $S_i$ and coincidence rates $C_{ij}, i,j \in \{A\Plus, A\Minus, B\Plus, B\Minus\}$ across different detectors can be estimated. 

$R_A$ is the ratio of transmissions $t_{A\Minus}$ over $ t_{A\Plus}$, where $t_{A\Minus}$ is the transmission through the input to output $A\Minus$ and $t_{A\Plus}$ is the transmission through the input to output $A\Minus$. $R_B$ is defined analogously. We measure values for $R_A$ and $R_B$ by sending a pulsed laser into the input and measuring the ratio of power transmitted across the output ports:

$$R_A = 0.99 \pm 0.03, ~~~~~~ R_B = 1.04 \pm 0.03$$

\begin{figure}[H]
    \centering
    \includegraphics[width=0.5\linewidth]{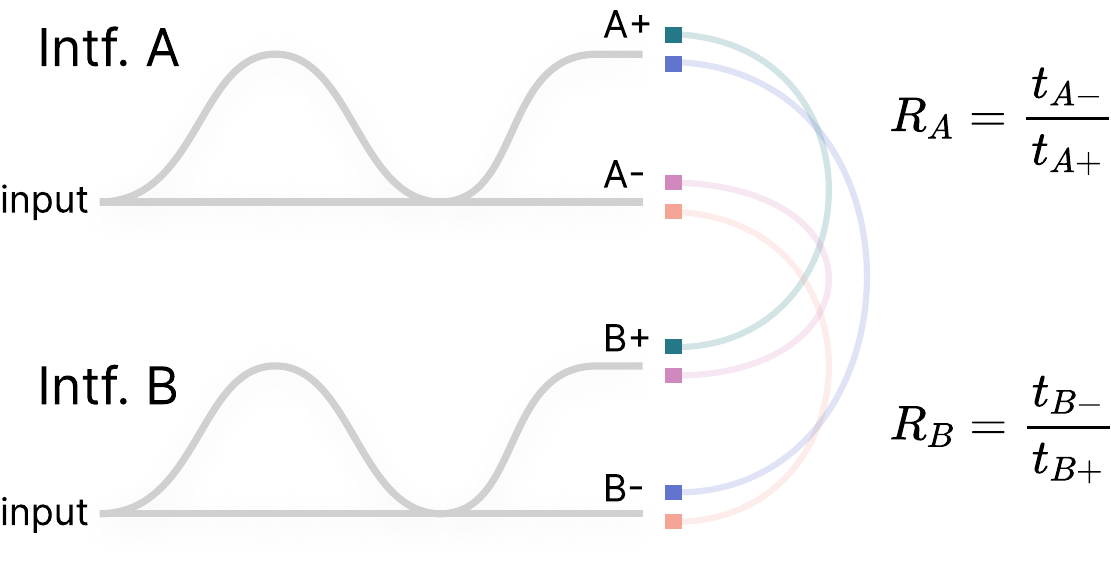}
\end{figure}

The coincidence rates for all wavefunction branches are the same if $\theta$ in \ref{cosines} is set to zero. With this phase state rate $C_{A\Plus B\Plus, \theta=0}$ measured directly (labeled B in \ref{fig:fringes}), it's straightforward to estimate the full wavefunction coincidence rate across all interferometer outputs. As shown in table \ref{table:rates}, the rate is $C_{A\Plus B\Plus}(1 + R_A + R_B + R_AR_B)$. This estimate is used to express a full coincidence rate for figures 3 and 4 in the main text. 
\\
\\
\begin{tabular}{ |p{2.5cm}||p{1.3cm}|p{2.9cm}|p{2.9cm}|p{3.4cm}|  }
 \hline
 \multicolumn{5}{|c|}{Predicted Singles and Coincidence Rates for Full Wavefunction} \\
 \hline
 rate metrics     &  Branch 1 & Branch 2 & Branch 3 & Branch 4 \label{table:rates}\\
 \hline
 singles rate 1  & $S_{A+}$ &$S_{A-}=R_A S_{A+}$    &$S_{A+}$               &$S_{A-}=R_A S_{A+}$        \\
 singles rate 2  & $S_{B+}$ &$S_{B+}$               &$S_{B-} = R_B S_{B+}$  &$S_{B-} = R_B S_{B+}$      \\
 coincidence rate&\textcolor{color1}{$C_{A+B+}$}&\textcolor{color3}{$C_{A-B+}=R_A C_{A+B+}$}&\textcolor{color2}{$C_{A+B-}=R_B C_{A+B+}$}&\textcolor{color4}{$C_{A-B-}=R_B R_A C_{A+B+}$}\\
 \hline
\end{tabular}
\\
\\
\\
For main-text figure 2 in particular, coincidences were measured with the center bin rate tuned to a minimum $\theta=-\frac{\pi}{2}$ (grey solid line in main-text figure 2c). The rate therefore captures the contribution from the 6 phase-independent terms for the $A\!\!+\!\!B\!+\!$ wavefunction branch. This can be extrapolated to the $\theta=0$ state by multiplying by $\frac{4}{3}$. For the rate displayed to be consistent with the measures of coupling efficiency $\eta$ (and so the interferometers aren't thought of as a source of $\sim$50\% loss), this rate is multiplied again by $(1 + R_AR_B)$ to represent two wavefunction branches. This results in a total scaling of: 

$$C_{fig. 2} = \frac{4}{3}(1 + R_AR_B)C_{measured}$$

\section{Time-Walk Correction}
SNSPD jitter increases with count rate due to properties of the nanowire reset process and features of the readout circuit. A threshold timing measurement set at a specific trigger level will 'walk' along the rising edge of SNSPD pulses by varying amounts if those pulses vary in amplitude and slew rate. At low count rates, SNSPDs exhibit very uniform pulse heights and shapes. However, at high counts rates where the inter-arrival time is on the order of the reset time of the detector, current-reset and amplifier effects lead to smaller and distorted pulses in a form of pulse 'pile-up'~\cite{Mueller2023}. 

\begin{figure}[H]
    \centering
    \includegraphics[width=0.7\linewidth]{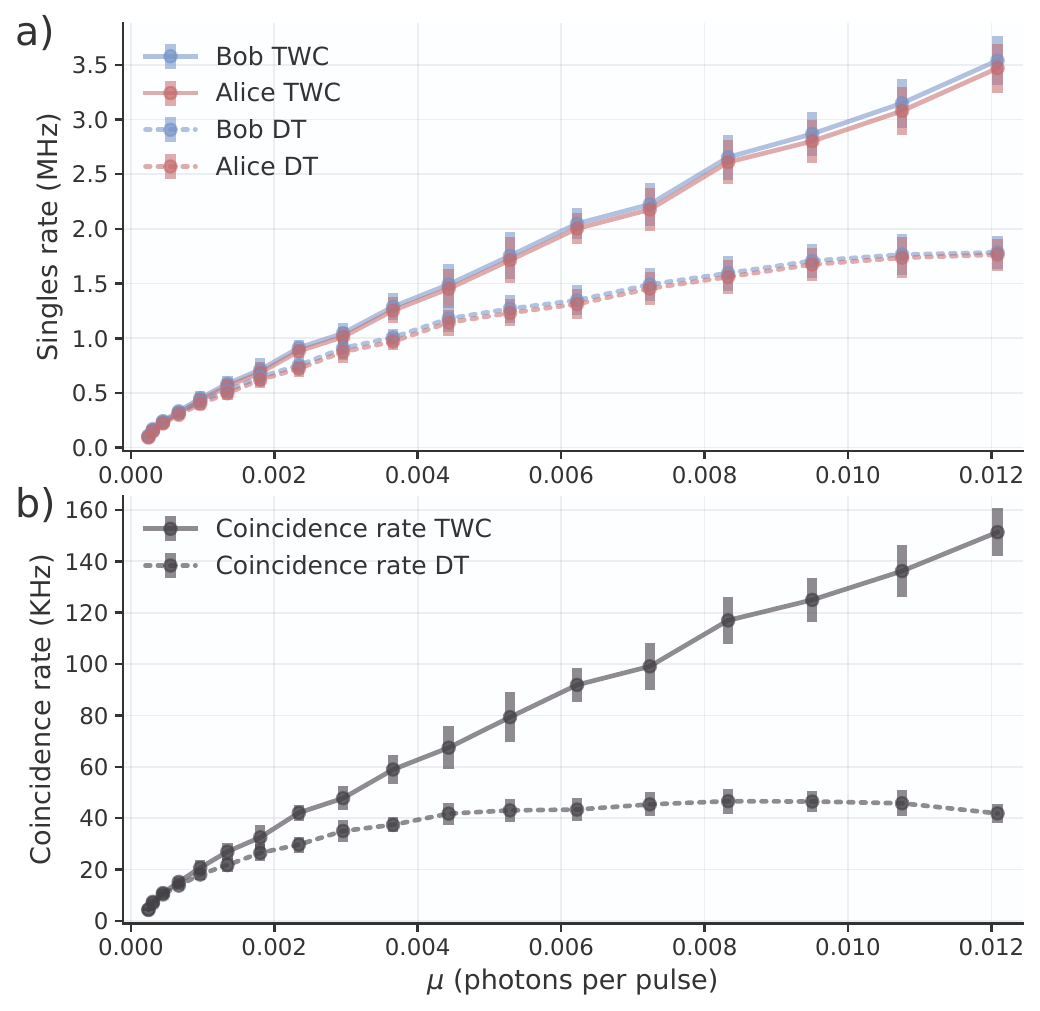}
    \caption{a) Adding 200~ns dead time (DT) to avoid high-rate jitter significantly reduces count rates at higher $\mu$. With time walk correction (TWC), the full available count rate is preserved while keeping the timing jitter low. b) The reduction in count rate for coincidences is more dramatic, because the lost efficiency at both detectors contributes. }
    \label{fig:time_walk_vs_dead_time}
\end{figure}

The effects of time walk can be filtered out by imposing a dead time following each detection. The length of the dead time is tuned so that SNSPD events arriving within it are expected to have distorted timing, and are thrown out. However, as shown in \ref{fig:time_walk_vs_dead_time}, this method can severely limit count and coincidence rates for detector types that exhibit long periods of undershoot or ringing on the falling edge of the RF pulse. For the differential SNSPDs used here, a 200~ns dead time would be necessary to filter out all time-walk effects. 

As detailed in \cite{Mueller2023, Craiciu23}, a calibration and correction process may be used to cancel out the effects of time walk without losing count rate. It relies on adding timing corrections $\tilde{d}$ to time-distorted SNSPD time tags based on the inter-arrival $t'$ time that precedes each tag. Building a lookup table for $\tilde{d}$ as a function of $t'$ is the objective of the calibration process. With this, real-time running software may largely cancel the increase in jitter from time-walk. 

Here we introduce a simplification of the calibration process from \cite{Mueller2023} that makes use of the pulse sequence already impinging on the detectors from the entanglement source. With the calibration routine written directly into our coincidence analysis software, recomputing the ideal $t'-vs-\tilde{d}$ curve takes no more than a few seconds whenever the detector bias currents or trigger levels are changed. 

We have previously shown how the $t'-vs-\tilde{d}$ curve can be found by illuminating the detector with photons from a pulsed laser source \cite{Mueller2023}. We imposed the requirement that the period $p$ of the pulses sequence be larger that the largest expected jitter observed due to time walk. This way, any time-tag can be unambiguously associated with the timing of the laser pulse that created it, thereby inferring the relation between $t'$ and $\tilde{d}$ for that event. 

Here we rely on a 2D histogram that plots $t'$ on the y-axis and the usual clock-referenced arrival time on the x-axis. Here, any single $t'$ and absolute arrival time measurement does not imply a correction $\tilde{d}$ when considered in isolation. This is because the time-distribution is an irregular pattern of histogram peaks of varying height, as opposed to a more uniform sequence. Also, delays $\tilde{d}$ can surpass the experiment's fundamental period (244.5~ps from the laser's 4.09~GHz rep rate), initially complicating the matching of trigger events to originating photon timing. But despite these nuances, the 2D histogram structure implies a method for extracting  the $t'-vs-\tilde{d}$ calibration curve through a special analyses. 

\begin{figure}[H]
    \centering
    \includegraphics[width=0.8\linewidth]{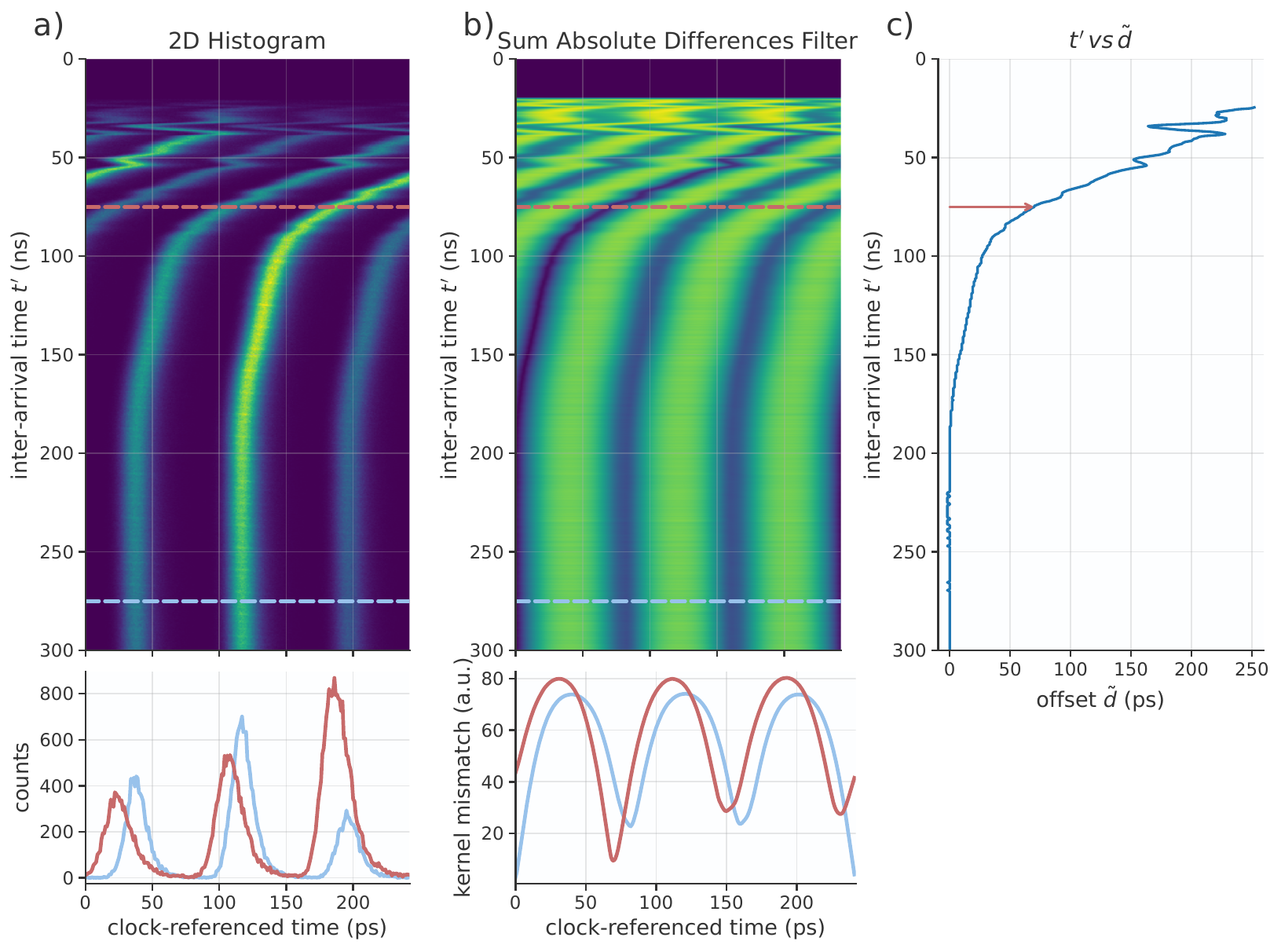}
    \caption{a) The 2d histogram with clock-referenced histograms on the x-axis and inter-pulse arrival time $t'$ on the y-axis. The effect of time walk is evident in the horizontal translations of the 3-peak structure as $t'$ decreases. Slices of the 2D histogram indicated by the red and blue horizontal line are plotted as regular histograms in the lower figure. b) After applying the Sum Absolute Differences filter to horizontal slices of the 2D histogram. The $t'-vs-\tilde{d}$ delay curve (c) is extracted from this by using the index of each row with minimum mismatch value as the $\tilde{d}$ value for that row.}
    \label{fig:time_walk}
\end{figure}

Slices at the bottom of the 2D histogram (fig. \ref{fig:time_walk}a) for large $t'$ are essentially identical to the low-count-rate 3-peak singles histograms like from figure 2c in the main text. As $t'$ decreases from here, the slice as a whole develops some linear offset or rollover (periodic every 244.5~ps). This offset is the $\tilde{d}$ offset of interest. Therefore, $t'-vs-\tilde{d}$ may be extracted by running a template matching algorithm on each horizontal slice, using the large-$t'$ slice as the template. We opt for a absolute differences algorithm (fig. \ref{fig:time_walk}b). As the algorithm progress to smaller and smaller $t'$, the offset may approach the fundamental period length and 'roll over' causing a jump-discontinuity. But this can be detected and corrected for, meaning the method may produce a $t'-vs-\tilde{d}$ curve (fig. \ref{fig:time_walk}c) with some $\tilde{d}$ larger than the fundamental period.

Overall, time-walk analysis based on the 2D histogram construction is straightforward to implement and user friendly because it may be applied \textit{in situ} as part of software that already creates histograms and records coincidences.

\section{16-Channel Configuration and Filter Bandwidths}
\begin{figure}[H]
    \centering
    \includegraphics[width=0.7\linewidth]{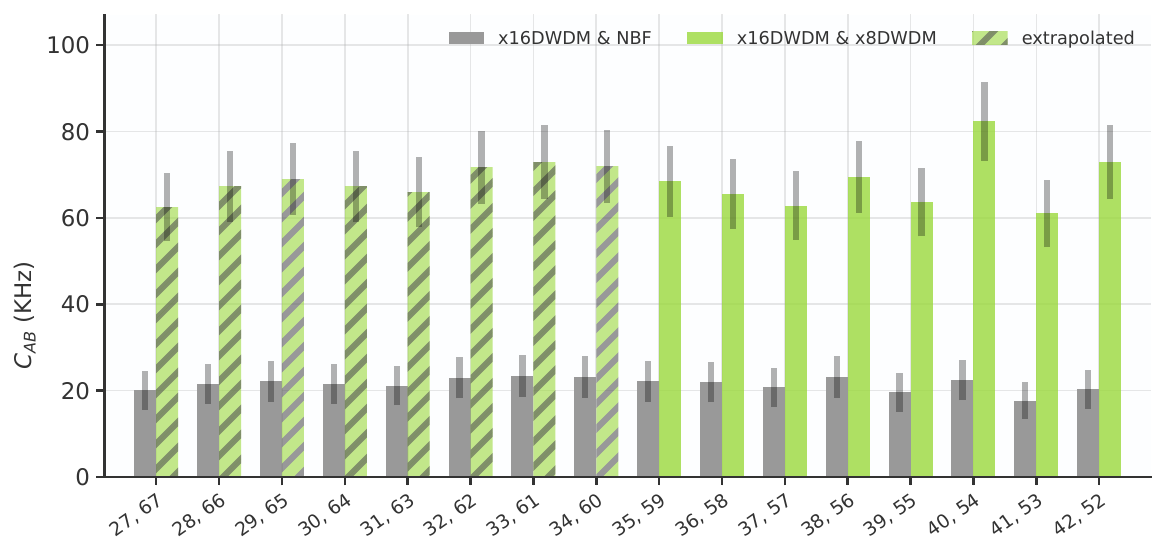}
    \caption{Coincidence rates for energy-matched channel pairings. The light green bars depict the main diagonal in (a). Grey bars are measured with x16 DWDM at Alice and a tunable narrowband filter at Bob. Dashed bars predict the rates for a system with x16 DWDMs at both Alice and Bob}
    \label{fig:16ch_config}
\end{figure}

\begin{figure}[H]
    \centering
    \includegraphics[width=0.7\linewidth]{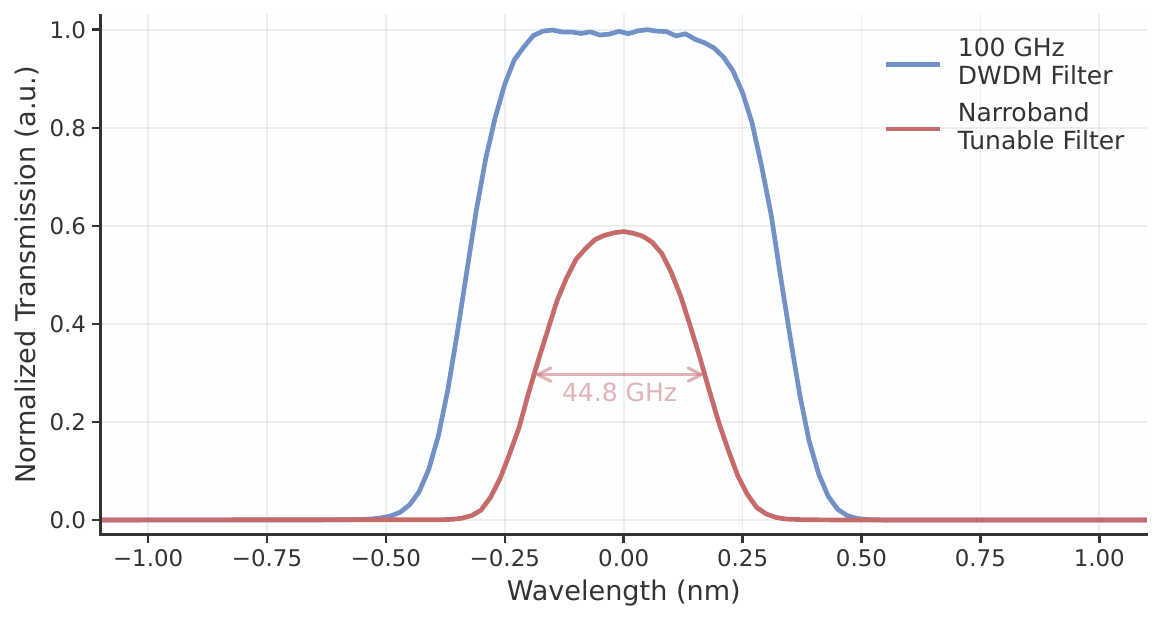}
    \caption{Spectrums for a single DWDM channel and narroband filter, with the center wavelengths set to zero. The narroband filter is used at Bob with the 16-channel DWDM at Alice to measure coincidence rates in the main text Fig. 2b. For the 100~GHz spec. DWDM filters, 100~GHz refers to the filter spacing. The FWHM passband for each is about 82~GHz.}
    \label{fig:filter_comparison}
\end{figure}

The energy-entangled spectrum of the SPDC extends beyond the bipartite spectrum spanned by the 8 DWDM channels. Therefore, we expect the source would offer higher total coincidence rate if more than 8 channel pairs are used. We investigate coincidence rates across 16 pairs by using all 16 channels available on the DWDM at Alice (24 - 34) and a tunable narrowband filter in place of the DWDM at Bob. As the narrowband filter has higher loss and 45~GHz FWHM passband (Fig.~\ref{fig:filter_comparison}), the coincidence rates are lower (grey bars in Fig.~\ref{fig:16ch_config}). But the uniformity of coincidence rates across 16 channels implies that the use of 16-channel DWDMs at both Alice and Bob would roughly double the total coincidence rate. The 8-channel DWDM has slightly lower loss than the 16-channel version. The impact of this discrepancy is negligible in the final system, though it played a role in choosing to investigate performance with both 16 and 8 channel versions. This is why one of each type was procured and ultimately used, instead of a symmetric 8-channel or 16-channel configuration.

\section{Guard Regions}
\begin{figure}[h]
    \centering
    \includegraphics[width=1\linewidth]{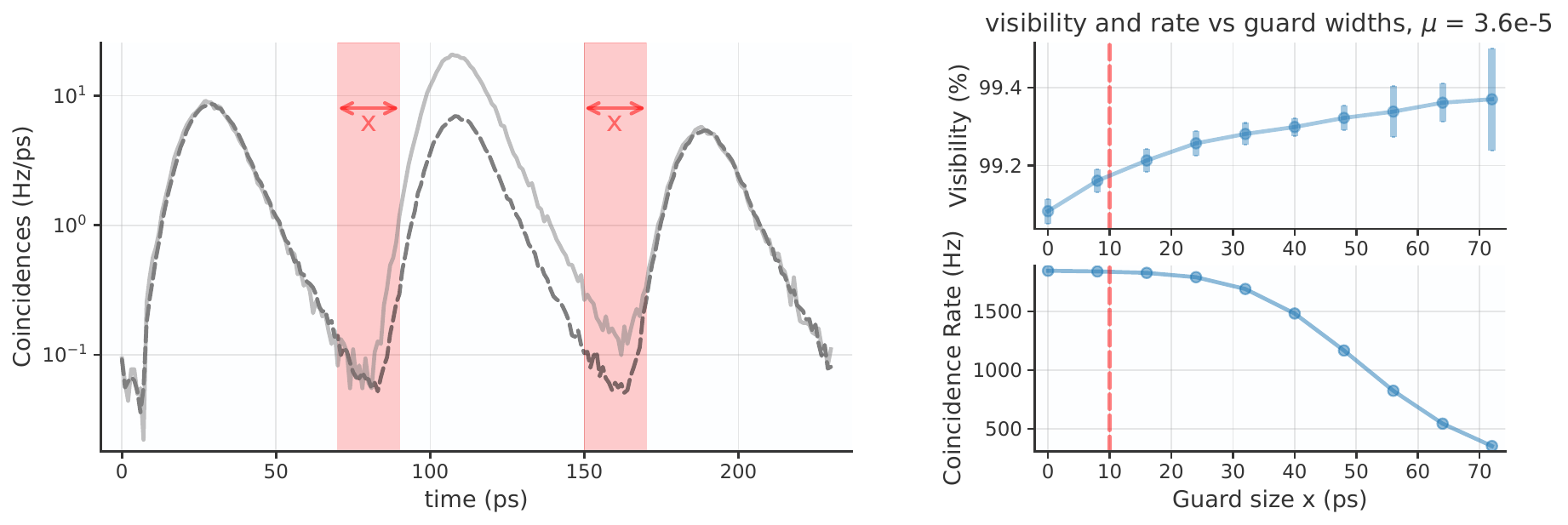}
    \caption{Charts on the right show phase basis fidelities and total coincidence rates as a function of width $x$, where both guard regions stay centered at 80 \& 160~ps.}
    \label{fig:guard_scan}
\end{figure}

The width $x$ of the guard regions centered at 80 and 160 ps were set to 10~ps. As shown in Fig. \ref{fig:guard_scan}, this width increase fidelities slightly without significantly impacting coincidence rates. 10~ps was not chosen based on rigorous analysis, though it would be possible to optimize the width to maximize some metric, like secret key rate. 

\section{Joint Spectral Intensity Analysis}

This entanglement source is potentially useful for future quantum communication systems that make use of two-photon interference, such as teleportation or entanglement swapping. These require the generation of highly pure and indistinguishable photons. The type-0 SPDC crystal used in this source shows strong spectral correlations that make it inappropriate for two-photon interference demonstrations by itself. However, with the application of the DWDM filtering, we create a series of spectral channel pairings that are potentially good individual sources of fairly pure single mode photons. To investigate this, we model the joint spectral intensity function (JSI) produced by the SPDC and fit it to data. Then we and apply a Schmidt decomposition to the spectral modes produced when pairs of DWDM filter channels are applied to the modeled JSI. We derive an inverse Shmidt number which is equal to single photon purity $P$ as well as visibility $V$ of Hong-Ou-Mandel (HOM) interference. An inverse Schmidt number approaching 1 indicates that the source is effective for two-photon interference measurements like HOM and Bell State measurements. 

\subsection{JSI modelling}

We follow an analysis for co-linear quasi-phase matching inside a waveguide packaged SPDC crystal \cite{Davis2022, ZielnickiKwiat2018SPDCmodel}. The joint spectral intensity $|f(\omega_s, \omega_i)|^2$  is modelled as a product of phase matching and pump envelope intensities |$\psi_{\mathrm{ph}}\left(\omega_s, \omega_i\right)|^2$ and $|\psi_p\left(\omega_s, \omega_i\right)|^2$, where $\omega_s$ and $\omega_i$ are the frequencies of the signal and idler models respectively. The phase matching envelope intensity takes the form:
$$\left|\psi_{\mathrm{ph}}\left(\omega_s, \omega_i\right)\right|^2=\operatorname{sinc}^2\left(\frac{\Delta k L}{2}\right),\;\;\;\;\;\; \Delta k=2 \pi\left(\frac{n\left(\lambda_p\right)}{\lambda_p}-\frac{n\left(\lambda_s\right)}{\lambda_s}-\frac{n\left(\lambda_i\right)}{\lambda_i}-\Gamma\right)$$
where $L$ is the length of the crystal, $\Delta K$ is a wave-vector mismatch term, and $n_{p(s)(i)}$ is the refractive index of the crystal at the wavelengths of pump $\lambda_p$, signal $\lambda_s$ and idler $\lambda_i$. $\Gamma = 1/\Lambda$ where $\Lambda$ is the polling period of the crystal. The refractive indices $n(\{\lambda_p, \lambda_s, \lambda_i\})$ are computed from an MgO-doped PPLN Sellmeier equation \cite{Gayer2008}. In the expression for $\Delta K$, $\lambda_p$ is the function of $\lambda_s$ and $\lambda_i$ that imposes energy conservation: $1/\lambda_p = 1/\lambda_s + 1/\lambda_i$. 

The pump envelope intensity is modeled as
$$|\psi_p\left(\omega_s, \omega_i\right)|^2=\exp \left(-\frac{\left(\omega_p-\omega_s-\omega_i\right)^2}{\sigma_p^2}\right)$$ where $\omega_p$ and $\sigma_p$ are the center frequency and bandwidth of the pump signal out of the EDFA/SHG module. Unlike $\lambda_p$ in the phase matching expression above, $\omega_p$ and $\sigma_p$ are fixed to known values or chosen as floating fitting parameters. 

\subsection{JSI fitting}
The JSI model is fitted to the the 8x8 DWDM data from Fig.~2a in the main text. The transmission spectrum of one DWDM channel was measured and used to estimate the transmission properties for all pairs of 8 channels. For each measurement of coincidence rate from Fig.~2a we define an integration over the joint spectral intensity and the corresponding filter passbands:

\begin{align}
    C_{u,v} = \int_{s}\int_{i}T_u(\lambda_s)T_v(\lambda_i)|f(\omega_s, \omega_i)|^2 d\lambda_s d\lambda_i \label{C_uv}
\end{align}

where $T_{u(v)}$ is the transmission spectrum for filter with ITU channel $u (v)$, and the integrations are over the signal and idler wavelengths. Efficiency parameters $\eta_i$ scale the DWDM filter spectrums, and are used to model all loss between pair generation in the SPDC crystal and detection in the SNSPDs. To fit the $\eta_i$ values, the singles rates from Fig.~2a were also included, and fitted to single-filter integrations:

$$S_{u} = \int_{s}\int_{i}T_{u}(\lambda_a)|f(\omega_s, \omega_i)|^2 d\lambda_s d\lambda_i$$

Where $a \in \{i, s\}$. Parameters of model for $|f(\omega_s, \omega_i)|^2$ were optimized to minimize the error between estimates $C_{u,v}$ and $S_u$ and the measured coincidence and singles rates. The following parameters were either set explicitly based on measurements and known constants (black), or optimized in the fitting process (blue):

\begin{figure}[H]
    \centering
    \includegraphics[width=1\linewidth]{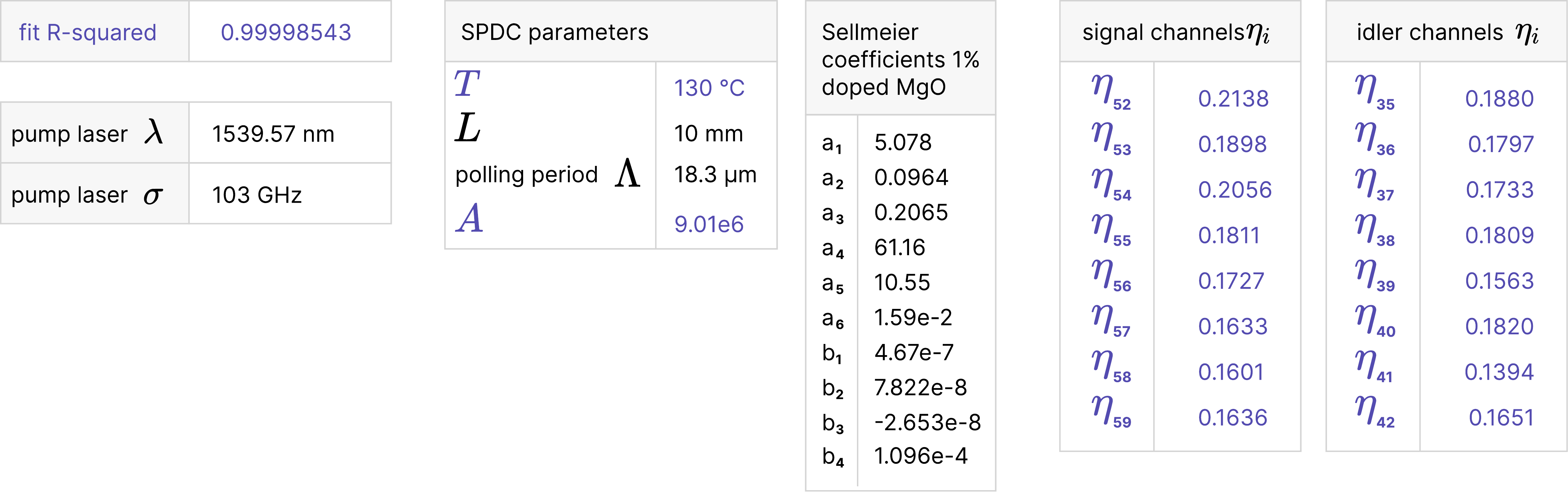}
    \caption{Pump laser $\lambda$ and $sigma$ were found by measuring the output of the SHG module with a spectrum analyzer. }
    \label{fig:parameters}
\end{figure}

Varying MgO doping percentage and varying crystal temperature affect the Sellmeier equations in similar ways \cite{Gayer2008, Jundt1997}. Therefore, crystal temperature $T$ was used as a fitting parameter, to stand in the unknown MgO doping percentage. The true crystal temperature during all measurements was $50.0^{\circ}~\mathrm{C}$, which was optimized to maximize coincidence rates.


\begin{figure}[H]
    \centering
    \includegraphics[width=0.9\linewidth]{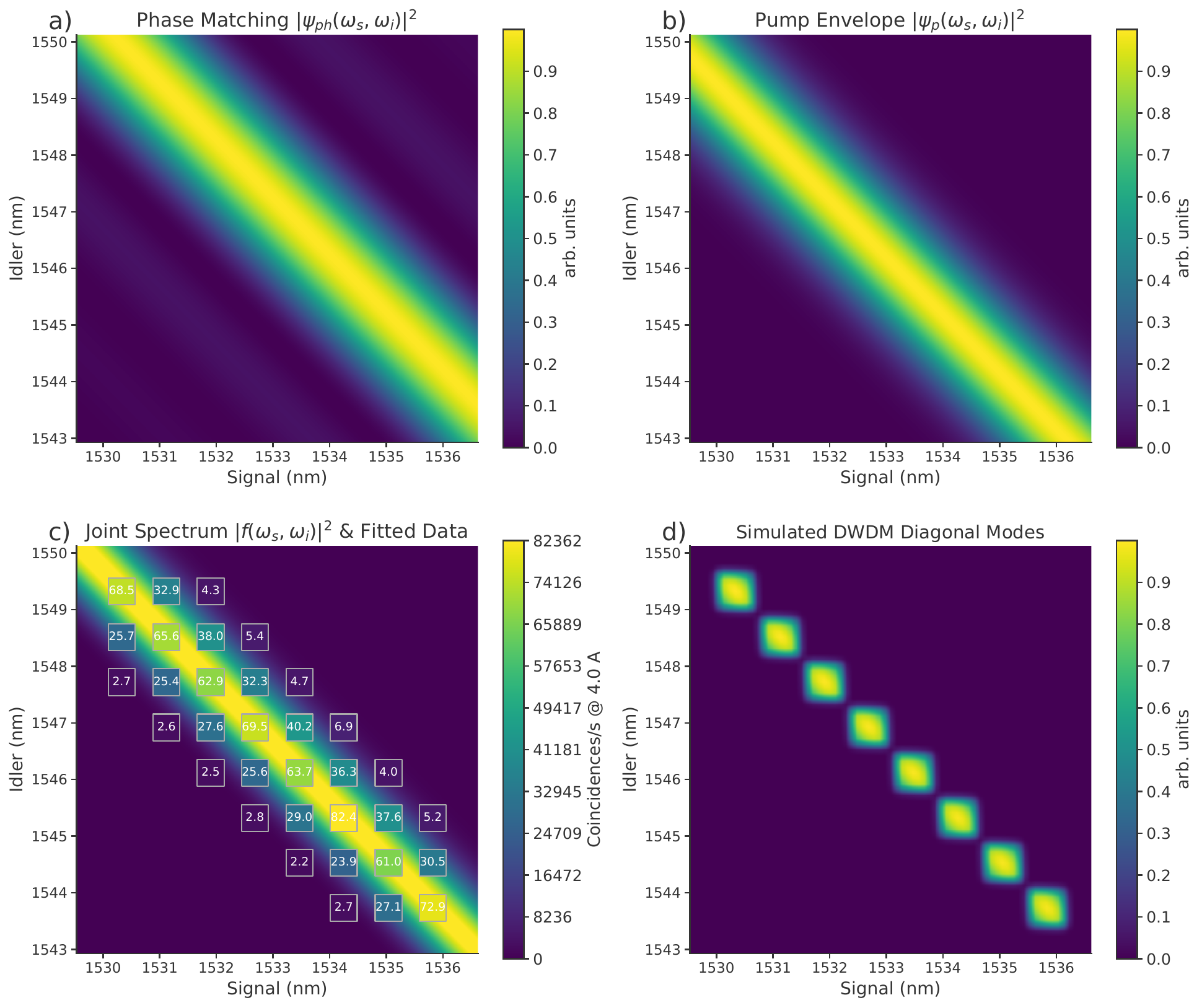}
    \caption{a) \& b) For type-0 SPDC, the relationship between refractive indices of pump and down-converted wavelengths produces an intensity profile roughly parallel with the pump envelope. This is responsible for the broad output spectrum of these crystals. c) the coincidence rates in Kcounts/s that are fit to filter-defined integrations of the JSI. d) Joint spectral distributions of coincidences expected from the filter pairings that correspond to the main diagonal. Here the joint spectrums of 8 DWDM channels pairs are shown summed together, where each component of the sum is an expression like the integrand of $C_{u,v}$ above. }
    \label{fig:jsi_sim}
\end{figure}

An ideal photon pair source for scalable optical quantum information processing would not exhibit joint spectral correlations between the signal and idler photons. The Schmidt decomposition, which is equivalent to the singular value decomposition in our computational model, can be used to quantify these correlations \cite{ZielnickiKwiat2018SPDCmodel}. We apply the Schmidt decomposition to the filtered JSIs for filter pairings along the main diagonal in Fig.~2a in the main text, and derive the inverse Schmidt number for these modes
$$1/K = \sum_i \lambda_i^2$$
where $\lambda_i$ are the Schmidt coefficients \cite{ZielnickiKwiat2018SPDCmodel}. The inverse Schmidt numbers for all 8 channel pairs are similar, and are not expected for vary due to any phenomena beyond inaccuracies of the model. Therefore, we quote single values for $1/K$ in the main text. 

\section{Consequences of Narroband Filtering}
\subsection{Calculation of mean photon number per pulse \texorpdfstring{$\mu$}{(mu)}}

\begin{figure}[H]
    \centering
    \includegraphics[width=1\linewidth]{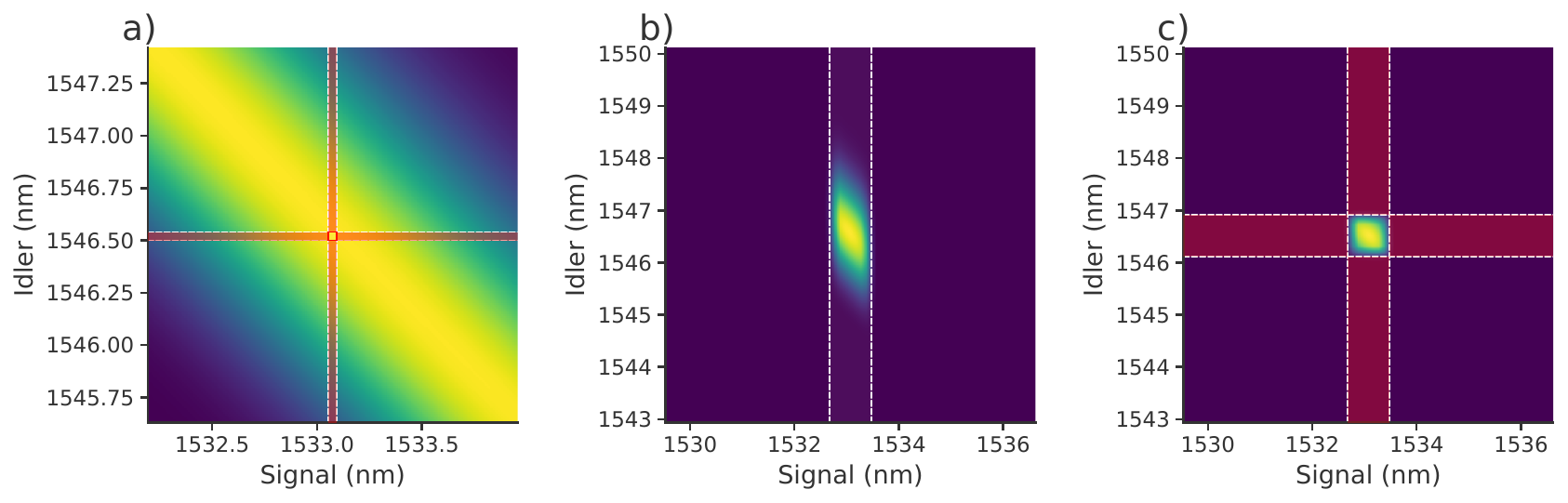}
    \caption{a) Two filters with bandwidth approximately 0.04 nm are applied to the JSI model. True coincidences are expected only from the region inside the small red square. b) The shape of the JSI with one 100 GHz DWDM filter applied c) the JSI a 100 GHz DWDM filter applied on both signal and idler arms. }
    \label{fig:narroband}
\end{figure}

It is common in the literature to calculate signal and idler arm efficiencies with equations of the form:
\begin{align}
\eta_{s(i)}=\frac{C_{s i}}{S_{i(s)}} \label{regular_eff}
\end{align}
Where $\eta_{s(i)}$ is coupling efficiency of the signal (idler) arm, $C_{s i}$ is the coincidence rate, and $S_{i(s)}$ is the singles rate on the idler (signal) arm. $\mu$, the mean pair rate per period or cycle, may also be defined in terms of the repetition rate $R$ as $R \mu \eta_{s(i)}=S_{s(i)}$. With these, one can define $\mu$ in terms of coincidence rate $C_{s i}$ and singles rates $S_s$ and $S_i$:
\begin{align}
    \mu=\frac{S_s S_i}{C_{s i} R} \label{eq:colorless}
\end{align}
However, this definition breaks down in the limit of narroband filtering, or when the losses on the signal and ider arm cannot be thought of as 'colorless'. Consider the situation of two very-narroband filters, as illustrated in figure \ref{fig:narroband}a. This situation can be simulated using the JSI model. We set signal and idler filter bandwidth to 5\% of the 100 GHz DWDM bandwidths. Pump power is scaled by 200x. Transmission of the filters at their maximum is set to 100\%. This results in $S_s = 58.0~\mathrm{MHz}$, $S_i = 57.6~\mathrm{MHz}$, and $C_{s i} = 567~\mathrm{KHz}$. With these values, eq. \ref{eq:colorless} suggests a $\mu$ value of 1.47. This is misleading because it's unreasonable to expect the red outlined region at the intersection of the filters in \ref{fig:narroband}a --which is responsible for all true detections of entangled pairs-- to be the source of more than one entangled pair per pulse. Rather, the high singles rates $S_s$ and $S_i$ are having an adverse effect on the $\mu$ calculation. Most of the singles detections are from mutually incompatible spectral modes -- the 4 regions that form a cross shape above, below, and to the sides of the red outlined square. 

We propose a definition of $\mu$ similar to the form of eq. \ref{C_uv}:
\begin{align}
    \mu_{u,v} = \frac{1}{R}\int_{s}\int_{i}W_u(\lambda_s)W_v(\lambda_i)|f(\omega_s, \omega_i)|^2 d\lambda_s d\lambda_i \label{newmu}
\end{align}
where $W_u$ and $W_v$ have the spectrums of filters $T_u$ and $T_v$ but maximum transmissions of 100\%. This is an integration of $|f(\omega_s, \omega_i)|^2$ over the bipartite spectral region where filter transmission is non-negligible. As the JSI model is defined, $|f(\omega_s, \omega_i)|^2$ has units of entangled pairs per $\mathrm{nm}^2$.

Going forward, we use the eq. \ref{newmu} definition of $\mu$ in the main text, and do a separate analysis of the effect of the mutually incompatible spectral modes when necessary. Note that in the main text, Alice receives idler photons and Bob receives signal photons, so variables transform as $C_{is} \rightarrow C_{AB}$, $S_{i} \rightarrow S_{A}$, and $S_{s} \rightarrow S_{B}$

\subsection{Estimating \texorpdfstring{$\mu$}{mu} from coincidence and singles rates}
As eq. \ref{eq:colorless} is problematic for the narroband filtering regime, a conversion is needed that maps this type of expression to the more implicitly defined \ref{newmu}. Luckily, the influence of narroband filtering as a sort of pseudo-loss can be rolled into a new geometric factor $\delta$. This is the ratio of the JSI captured by two narroband spectral filters $W_u(\lambda_s)*W_v(\lambda_i)$ to that captured by one $W_v(\lambda_i)$, as illustrated in \ref{fig:narroband}b and c. Consider an expression for $\delta$ in terms of measured and fitted quantities:
\begin{align} 
\delta_{i(s)} = \frac{C_{is}}{\eta_{i(s)}S_{s(i)}} \label{new_eff}
\end{align}
Here, $S_{s(i)}$ is the singles rate through a signal (idler) narrowband filter. $\eta_{i(s)}S_{s(i)}$ would be the coincidence rate if the idler (signal) filter was wide-band (or just colorless loss), but maintained the same efficiency as measured ($\eta_{i(s)}$). Therefore the fraction can be thought of as the ratio of a coincidence rate measurement with two narroband filters $[C_{is}]$ to a coincidence rate measurement with one narro-band and one wide-band filter $[\eta_{i(s)}S_{s(i)}]$. As we use the same bandwidth filters on the signal and idler arms, the $\delta_i$ and $\delta_s$ calculated from the measured data and JSI fitting results are similar. We use an averaged single $\delta$ value for all further analysis, unique to our JSI bandwidth and 100 GHz DWDM filters. 
$$\delta = 0.393 \pm 0.012$$
This is averaged from 8 $\delta_i$ and 8 $\delta_s$ values, for the 8 DWDM channel pairs along the main diagonal of the JSI. 

We see eq. \ref{new_eff} is a modified form of \ref{regular_eff} that includes $\delta$. With $R \mu \eta_{s(i)}=S_{s(i)}$ again, we can define a new expression for $\mu$ using singles $S_{s(i)}$ and coincidence $C_{is}$ rates. 
\begin{align}
    \mu = \frac{\delta S_s S_i }{R C_{is}} \label{new_mu}
\end{align}
This is used to compute $\mu$ in the main text. 




\section{Quantum State Tomography}

We perform quantum state tomography of individual channel pairs in order to resolve the full density matrix and calculate bounding terms for the distillable entanglement rate. 

\begin{figure}[H]
    \centering
    \includegraphics[width=0.8\linewidth]{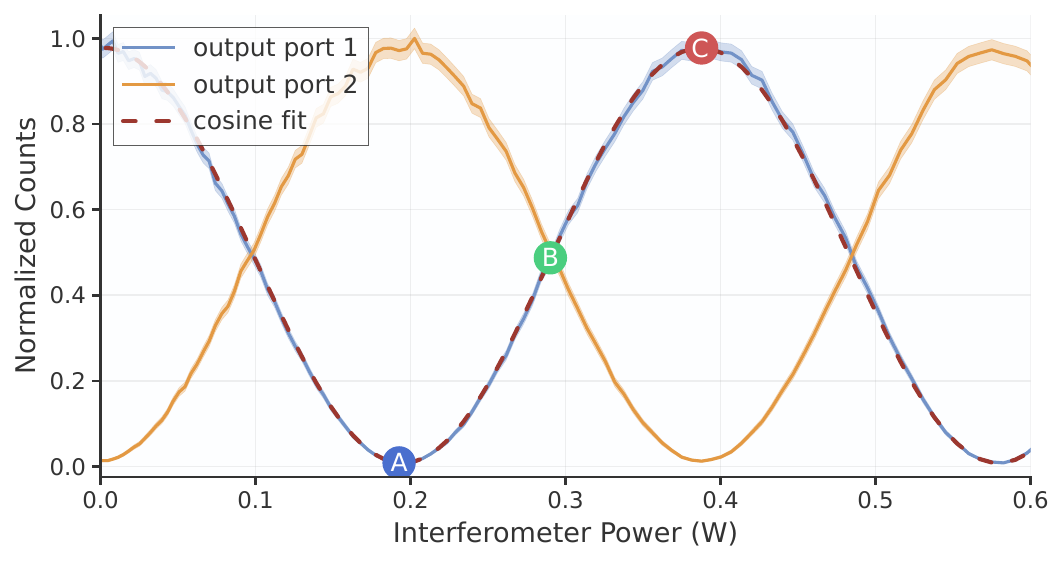}
    \caption{Coincidence rate across center time bin with respect to interferometer power. The fringe data was measured for the two output ports of Alice's interferometer. The output port 1 data was fitted to a cosine curve using linear least squares.}
    \label{fig:fringes}
\end{figure}

Established methods for tomography of time-bin entangled photon pairs involve independent phase control of the two readout interferometers. However, the center bin coincidence rate depends on an overall phase which is the sum of source and readout interferometer phases (see \ref{cosines}). By manipulating the overall phase through control of only one interferometer, we assume that the full phase-dependent behavior of the system is captured. The data for the tomography is captured with Alice's interferometer in 3 phase settings. That which minimizes coincidence rate of the center bin (labelled A in \ref{fig:fringes}), that which maximizes coincidences in the center bin (C), and a state halfway between in phase (B). The (B) state is chosen by setting the interferometer power to a value halfway between the power settings for states A and C. We therefore assume phase scales linearly with power, which is supported by the good agreement in \ref{fig:fringes} between the phase fringes and a cosine fit.. 

Multiple measurements involving all three time bins are performed with the phase set in these 3 states. The columns labeled A, B, and C in \ref{fig:counts_chart} correspond to these 3 measurement conditions, where the measurements for the 90 degree phase measurements are repeated in columns 2 and 3. We organize the coincidence data in this format to show how it relates to more canonical constructions of tomography data where the phases of the two readout stations are controlled independently \cite{Zhang2021Tomography,Takesue09Tomography}. Summing the counts in each row yeilds a 16-element vector which is sufficient to calculate a density matrix using a maximum likelyhood method \cite{tomography2023}. 

\begin{figure}[H]
    \centering
    \includegraphics[width=1\linewidth]{Chart_1.pdf}
    \caption{Quantum state tomography data for one low-power (SHG at 1.2 Amps) measurement of channels 35 and 59. The numbers are coincidences per second. Dashes (-) indicate projections that cannot be measured for the particular pairing of time-bin choice (row) and interferometer phase setting (column).}
    \label{fig:counts_chart}
\end{figure}

We use a density matrix to calculate bounds on $E_D$, the distillable entanglement rate. $E_D$ quantifies the rate of maximally entangled Bell pairs which may be created from the received state with local operations and classical communication. $E_D$ is bounded above by log-negativity $E_N$ and below by coherent information ($E_I = \mathrm{max}\{I_{A\rightarrow B}$, $I_{B\rightarrow A}\}$) \cite{Alshowkan2022,Eisert2000} both of which may be calculated from the density matrix:

$$E_N = \mathrm{log_2}||\rho^{A}|| \quad\quad\quad I_{A\rightarrow B} = H\left(\rho^\mathrm{B}\right)-H\left(\rho^{\mathrm{AB}}\right)$$

Where $||\rho^{T_A}||$ is the trace norm of the partial transpose of $\rho$, the calculated bipartite density matrix. $H$ is the base-2 von Neumann entropy \cite{Vidal2002negativity, Devetak2004coherent} 

The entanglement rate of the experiment in ebits/s is the coincidence rate $C_{AB}$ times $E_D$. We therefore plot $C_{AB}E_N = C_N$ and $C_{AB}E_I = C_I$ in the main text as the upper and lower bounds on entangled bit rate $C_{AB}E_D = C_D$. Multiple integrations are performed at each phase and power setting, so that multiple density matrices and multiple measures of coherent information and log-negativity may be derived. The averages and standard deviations of these sets of measurements are used to define the values and error bars for figures 3 and 4 in the main text. 

Figure 3 in the main text also shows a lower bound on secret key rate (SKR) given reasonable assumptions for a slightly different implementation of the entanglement source. This is calculated with a key genration rate formula \cite{ma2007quantum}:

\begin{align}
C_{SKR} &= C_{AB} E_S \\
C_{SKR} &= q C_{AB}[1-f(\mathcal{E}) H_2(\mathcal{E})-H_2(\mathcal{E})]
\end{align}
Where $E_S$ is the secret key fraction, $C_{AB}$ is the raw coincidence rate, $q$ is a basis reconciliation factor, $\mathcal{E}$ is the quantum bit error rate, and $f(x)$ is the bidirectional error correction efficiency. Quantum bit error rate $\mathcal{E}$ is $(1-V)/2$ where $V$ is visibility~\cite{Kim2022}. We choose a constant realistic value for $f(\mathcal{E})$ of 1.1 \cite{ElkoussReconciliation2010}. $H_2(x)$ is the binary entropy formula defined as
$$\mathrm{H}_2(x)=-x \log _2(x)-(1-x) \log _2(1-x).$$
We use a basis reconciliation factor of 0.81, thereby assuming a readout configuration at both Alice and Bob where 90\% of each channel's flux is directed to a z-basis measurement and may be used to build the shared key. The other 10\% is used for phase-basis measurements with the interferometers.

\begin{figure}[H]
    \centering
    \includegraphics[width=0.7\linewidth]{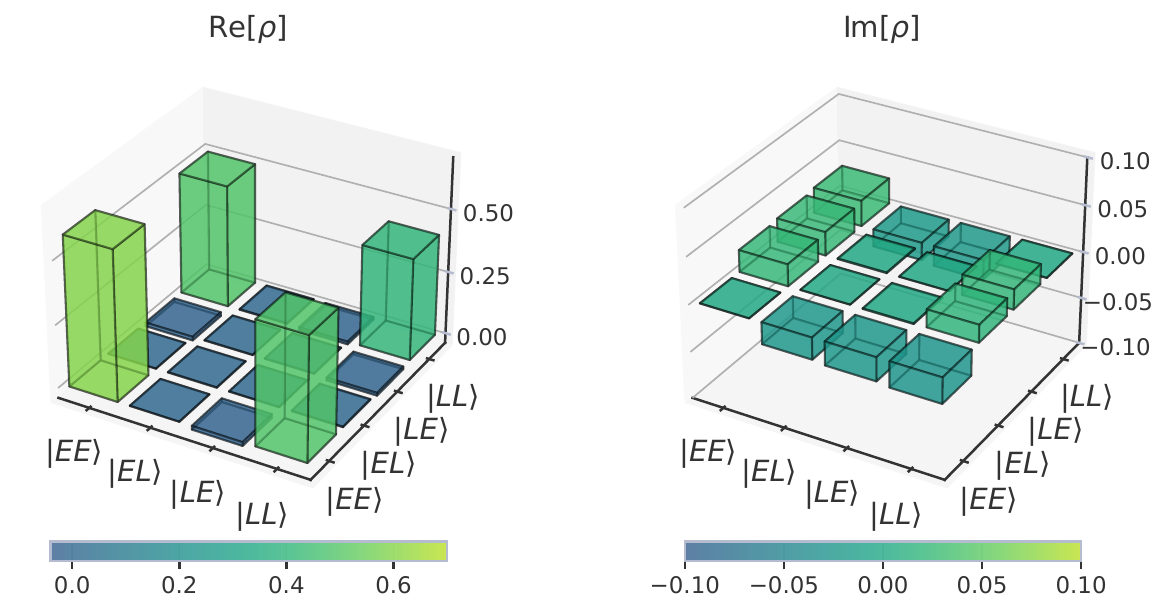}
    \caption{Density matrix data from channels 35 and 59 at 1.2 A SHG pump power. The $|ee\rangle$ state is higher than all other states due to the interferometer imbalances. For the matrix shown here, $E_N = 0.971$ and $I_{A\rightarrow B} = 0.904$}
    \label{fig:density_matrix}
\end{figure}

\section{Incompatible Bases \& Accidental Coincidence Rate}

An entangled pair can be called 'spectrally compatible' with a given DWDM channel pairing if --given no losses-- it would be detected 100\% of the time across that channel pair. In the case where signal and idler modes are perfectly spectrally compatible, it has been shown that accidental coincidences still negatively impact visibilities \cite{Kim2022}. In this case and assuming negligible dark counts, visibility is reduced as: 
\begin{align}
V = \frac{1}{1+\mu}V_0
\end{align}
Where $V_0$ is a nominal interferometer-limited visibility \cite{Kim2022} and $\mu$ is the mean photon number per pulse (classically defined like with \ref{eq:colorless}). The accidental coincidence rate discussed in the main text $C_{Acc}$ is only partially related to this, as it specifically takes into account only accidental coincidences due to incompatible spectral modes. Such coincidences can arise from two situations, and can be assigned their own coincidence rates
\begin{itemize}
\item $C_{ee}$: two photons both from mutually incompatible spectral regions, like the red regions in Fig. \ref{fig:narroband}c 
\item $C_{em}$ and $C_{me}$: one photon from the central overlapping filter region and one from an incompatible spectral region. 
\end{itemize}

Say one member of a spectrally compatible entanglemed pair reaches Alice, but not Bob due to loss. The photon received at Alice could form a coincidence with a spectrally incompatible photon that arrives at Bob. These are the $C_{em}$ and $C_{me}$ type coincidences. $C_{Acc}$ in the main text is the sum of $C_{ee}$, $C_{em}$ and $C_{me}$:

\begin{align}
    C_{Acc} &= C_{ee} + C_{em} + C_{me} \\
    C_{ee}/R &= (1 - \delta)\frac{S_A}{R} * (1 - \delta)\frac{S_B}{R} \\
    C_{em}/R &= (1 - \delta)\frac{S_A}{R} * \delta (1-\eta_A) \frac{S_B}{R} \\
    C_{me}/R &= (1 - \delta)\frac{S_B}{R} * \delta (1-\eta_B) \frac{S_A}{R} \\
\end{align}
\begin{align}
C_{Acc} = \frac{1}{R}(1 - \delta) S_A S_B \left(\delta \left(1-\eta _A\right)+\delta \left(1-\eta _B\right)+1-\delta\right)
\end{align}

\section{Impact of experimental imperfections on entanglement visibility}

\begin{figure}[h!]
    \centering
    \includegraphics[width=\textwidth]{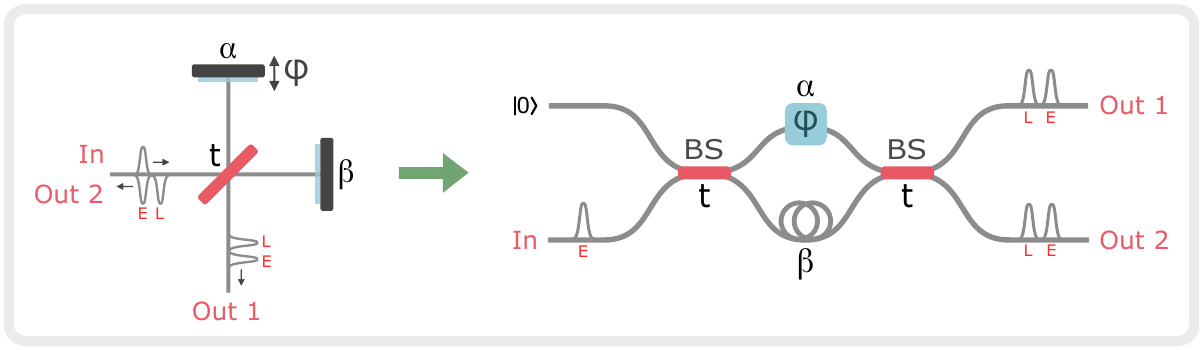}
    \caption{Model for Michaelson interferometers employed in the experiment. The interfomerter contains a beamsplitter with transmittance $t$ and two mirrors with efficiencies $\alpha$ and $\beta$.}
    \label{fig:interf_model}
\end{figure}

The experiment employs three Michaelson interferometers with a path-length delay of 80 ps: one at the source to generate the early and late time-bins, and one prior to each detector to control the measurement basis. To determine the effect of interferometric imperfections on the entanglement visibility, we model the interferometers as equivalent Mach-Zehnder interferometers as shown in Fig. \ref{fig:interf_model}. Imperfections in the interferometer are captured by the transmittance $t$ of the beamsplitter and internal path (mirror) efficiencies $|\alpha|^2$ and $|\beta|^2$. An ideal Michaelson interferometer has $t = 1/\sqrt{2}$ and $|\alpha|^2 = |\beta|^2  = 1$.

\subsection{Source Interferometer Imperfections}
In the experiment, pulses of coherent light from a mode-locked laser (MLL) are injected into the input of the source interferometer. A field $\hat{E}$ at the input of the source interferometer transforms as
\begin{align}
    \hat{E}_{\text{in}}  \rightarrow rt \alpha e^{i\varphi} \hat{E}_{E,1} +  r^2 \alpha e^{i\varphi}\hat{E}_{E,2} + rt\beta \hat{E}_{L,1} + t^2\beta \hat{E}_{L,2} + ir\sqrt{1-|\alpha|^2}\hat{E}_{\text{vac}_1} +it\sqrt{1-|\beta|^2}\hat{E}_{\text{vac}_2} \label{eq:source_field_transform}
\end{align}
where the early and late temporal modes are denoted by subscripts "E" and "L", the input and output modes are denoted by subscripts "in", "1" and "2", and $r = i\sqrt{1-|t|^2}$. Due to imperfect path efficiencies, part of the light leaks into the vacuum field mode $\hat{E}_{\text{vac}}$, which corresponds to the last term in Eq. \ref{eq:source_field_transform}. It follows that the power of the early and late output pulses in terms of the power of the input pulse are
\begin{align}
    &P_{E,1} = |r|^2|t|^2|\alpha|^2 P_{in},&P_{E,2} = |r|^4 |\alpha|^2 P_{in} \label{eq:powers}\\
    &P_{L,1} = |r|^2|t|^2|\beta|^2 P_{in} ,&P_{L,2} = |t|^4 |\beta|^2 P_{in}.\nonumber 
\end{align}



To generate the entangled photon pairs, one of the output ports of the source interferometer is up-converted by second harmonic generation (SHG)
then down-converted via spontaneous parametric down conversion (SPDC), resulting in two-mode squeezed vacuum states (TMSVs) in early and late temporal modes with mean photon numbers $\mu_E$ and $\mu_L$, respectively. The ratio of $\mu_E$ to $\mu_L$ depends on which output port of the source interferometer is used. Note that the definition of $\mu$ used in the main text is per source laser period or per experiment cycle (4.09 GHz). Therefore $\mu$ from the main text is equal to $\mu_E + \mu_L$. 
The output power of SHG ($P_{SHG}$) as a function of the SHG pump power ($P_p$) is \cite{parameswaran2002observation},
\begin{align}
    P_{SHG} = P_{p}\tanh^2{\sqrt{\eta_{SHG}P_{p}}} \approx \eta_{SHG}P_{p}^2
\end{align}
where $\eta_{SHG}$ is the conversion efficiency of the SHG crystal.
After SPDC, the squeezing parameter ($\xi$) of the TMSVs in terms of the SPDC pump power ($P_{SHG}$) is $\xi = \lambda \sqrt{P_{SHG}} \approx \lambda \sqrt{\eta_{SHG}}P_p$, where $\lambda$ is proportional to the SPDC crystal length and nonlinear interaction strength \cite{kaiser2016fully}. 
The mean photon number in terms of the squeezing parameter is $\mu = \sinh^2{\xi}\approx \xi^2$. Therefore, the mean photon numbers of the TMSVs as a function of the output pulses of the source interferometer are,
\begin{align}
    &\mu_E \approx \lambda^2 \eta_{SHG} P_{E,i}^2 &\mu_L\approx \lambda^2 \eta_{SHG} P_{L,i}^2
\end{align}
where $i = 1(2)$ corresponds to output port 1(2) of the source interferometer.

If output port 1 is used, $$\mu_E/\mu_L \approx P_{E,1}^2/P_{L,1}^2 = |\alpha|^4/|\beta|^4,$$ whereas if output port 2 is used,  $$\mu_E/\mu_L \approx P_{E,2}^2/P_{L,2}^2 = |r|^8|\alpha|^4/|t|^8|\beta|^4.$$
When output port 1 of the source interferometer is used, if the internal path efficiencies of the source interferometer are different, there is an imbalance in the early and late mean photon numbers. When output port 2 is used, the effect of the imbalance in the internal path efficiencies on the ratio of early and late mean photon numbers can be compensated by imperfect transmittance: $\mu_E/\mu_L = 1$ when $|t|^2/|r|^2 = |\alpha|/|\beta|$.

\subsection{Measurement Interferometer Imperfections}
\begin{figure}[h!]
    \centering
    \includegraphics[width=0.8\textwidth]{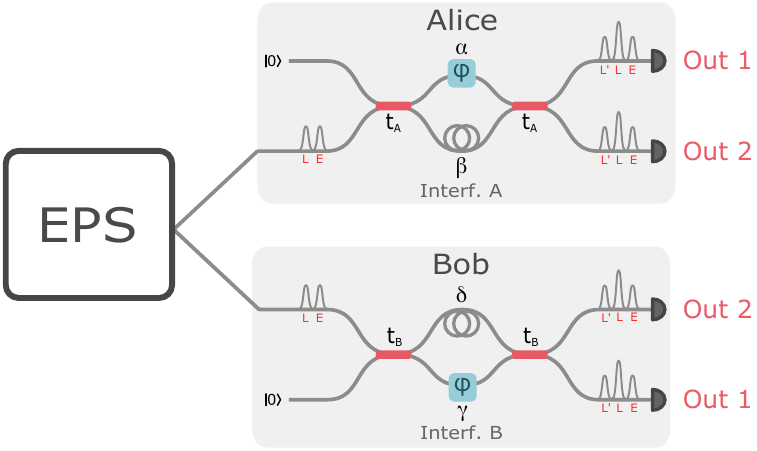}
    \caption{Setup for theoretical model of entanglement visibility experiment}
    \label{fig:ent_visib_model}
\end{figure}
 Imperfections in the measurement interferometers limit the entanglement visibility of the experiment. As described in the previous section, early and late TMSVs are generated by pumping the SPDC with early and late pulses. Each half of each TMSV is sent to a measurement interferometer (see Fig. \ref{fig:ent_visib_model}). Let $\ket{\xi}$ denote the TMSV state,
\begin{align}
    \ket{\xi} = \sum_{n=0}^{\infty}(-1)^n \sqrt{\frac{\mu^n}{(1+\mu)^{n+1}}}\ket{n_A,n_B}
\end{align}
where $\ket{n_A,n_B}$ denotes the state with $n_A$ photons at the input of interferometer A and $n_B$ photons at the input port of interferometer B. 
We model the input state to the measurement interferometers as a product state of TMSV in early and late temporal modes, to lowest order in $\mu_E$ and $\mu_L$:


\begin{align}
    \ket{\Psi_{in}} =& \ket{\xi}_E\otimes \ket{\xi}_L \nonumber \\
    = &\left(\frac{1}{\sqrt{1+\mu_E}}\ket{0,0}_E -\sqrt{\frac{\mu_E}{(1+\mu_E)^2}}\ket{1,1}_E +\cdots\right)
    \otimes \left(\frac{1}{\sqrt{1+\mu_L}}\ket{0,0}_L -\sqrt{\frac{\mu_L}{(1+\mu_L)^2}}\ket{1,1}_L +\cdots\right)\nonumber\\
    \approx& \sqrt{1-(\mu_E +\mu_L)}\ket{0,0}_E \ket{0,0}_L -\sqrt{\mu_E}\ket{1,1}_E  \ket{0,0}_L  -\sqrt{\mu_L}\ket{0,0}_E  \ket{1,1}_L\label{eq:input_state}
\end{align}

We can express Eq. \ref{eq:input_state} in terms of the creation operators  $\hat{a}^\dagger$ and $\hat{b}^\dagger$ of the field modes at the inputs of interferometers A and B, respectively:
\begin{align}
    \ket{\Psi_{in}} =\Big(\sqrt{1-(\mu_E+\mu_L)}-\sqrt{\mu_E} \hat{a}_{E}^{\dagger}\hat{b}_{E}^{\dagger} - \sqrt{\mu_L} \hat{a}_{L}^{\dagger}\hat{b}_{L}^{\dagger} \Big)\ket{0,0}_E\ket{0,0}_L \label{eq:input_state_fields}
\end{align}

Since the measurement interferometers are also Michaelson interferometers, the transformation relations are,
\begin{align}
    &\hat{a}_{E} \mapsto r_At_A \alpha e^{i\varphi} \hat{a}_{E,1} + r_A^2 \alpha e^{i\varphi}\hat{a}_{E,2} + r_At_A\beta \hat{a}_{L,1} + t_A^2\beta \hat{a}_{L,2}+c_A\hat{a}_{\text{vac}_1}+d_A\hat{a}_{\text{vac}_2},\label{eq:aE_transform} 
   \\
    &\hat{b}_{E}  \mapsto r_Bt_B \gamma e^{i\varphi} \hat{b}_{E,1} + r_B^2 \gamma e^{i\varphi}\hat{b}_{E,2} + r_Bt_B\delta \hat{b}_{L,1} + t_B^2\delta \hat{b}_{L,2}+c_B\hat{b}_{\text{vac}_1}+d_B\hat{b}_{\text{vac}_2},\label{eq:bE_transform}\\
    &\hat{a}_{L} \mapsto  r_A t_A \alpha e^{i\varphi} \hat{a}_{L,1}  + r_A^2 \alpha e^{i\varphi}\hat{a}_{L,2} + r_A t_A\beta \hat{a}_{L^\prime,1}+ t_A^2\beta \hat{a}_{L^\prime,2}+c_A\hat{a}_{\text{vac}_1}+d_A\hat{a}_{\text{vac}_2},\label{eq:aL_transform}\\
    &\hat{b}_{L}  \mapsto r_B t_B \gamma e^{i\varphi} \hat{b}_{L,1} + r_B^2 \gamma e^{i\varphi}\hat{b}_{L,2}  + r_B t_B\delta \hat{b}_{L^\prime,1}+ t_B^2\delta \hat{b}_{L^\prime,2}+c_B\hat{b}_{\text{vac}_1}+d_B\hat{b}_{\text{vac}_2},\label{eq:bL_transform}\\
    &c_A =ir_A\sqrt{1-|\alpha|^2}, \quad d_A=it_A\sqrt{1-|\beta|^2},\quad
    c_B =ir_B\sqrt{1-|\delta|^2},\quad\ d_B = it_B\sqrt{1-|\gamma|^2},\nonumber
\end{align}
where $L^\prime$ denotes the temporal mode obtained by sending a photon in the late ($L$) mode through the long arm of an interferometer, and $\hat{a}_{\text{vac}_i}$, $\hat{b}_{\text{vac}_i}$ correspond to vacuum modes. To find the state at the output of the interferometers, we combine Eq. \ref{eq:input_state_fields} with Eq. \ref{eq:aE_transform}-\ref{eq:bL_transform}, and consider only terms relevant to post-selection on coincidences of the middle bins (L) of different interferometer outputs, to lowest order in $\mu_E$, $\mu_L$,
\begin{align}
    \ket{\Psi_{\text{out}}}= 
    \cdots&+r_A^*t_A r_B^*t_B\Big(\beta \delta\sqrt{\mu_E}+\alpha \gamma \sqrt{\mu_L} e^{-2i\varphi}  \Big)\ket{0,0;0,0}_E\ket{1,0;1,0}_L\ket{0,0;0,0}_{L^\prime}\label{eq: output_state}\\
    &+r_A^*t_A\Big(t_B^2  \beta\delta \sqrt{\mu_E} + (r_B^*)^2  \alpha \gamma \sqrt{\mu_L} e^{-2i\varphi}\Big)\ket{0,0;0,0}_E\ket{1,0;0,1}_L\ket{0,0;0,0}_{L^\prime} \nonumber\\
    &+r_B^*t_B\Big(t_A^2\beta \delta\sqrt{\mu_E}  +(r_A^*)^2 \alpha \gamma  \sqrt{\mu_L} e^{-2i\varphi}\Big)\ket{0,0;0,0}_E\ket{0,1;1,0}_L\ket{0,0;0,0}_{L^\prime} \nonumber\\
    &+\Big(t_A^2 t_B^2\beta \delta\sqrt{\mu_E} + (r_A^*)^2 (r_B^*)^2 \alpha \gamma \sqrt{\mu_L} e^{-2i\varphi}\Big)\ket{0,0;0,0}_E\ket{0,1;0,1}_L\ket{0,0;0,0}_{L^\prime} + \cdots \nonumber 
\end{align}

where $\ket{n_{A,1},n_{A_2};n_{B_1}, n_{B,2}}$ denotes the state with $n_{A,1}$ photons at output 1 of interferometer A, $n_{A,2}$ photons at output 2 of interferometer A, $n_{B,1}$ photons at output 1 of interferometer B, and $n_{B,2}$ photons at output 2 of interferometer B. We define the following parameters to simplify notation:
\begin{align}
    x \equiv \frac{\mu_E}{\mu_L},\quad\quad\quad
    \kappa_A \equiv \frac{|\beta|^2}{|\alpha|^2}, \quad\quad\quad \kappa_B \equiv \frac{|\gamma|^2}{|\delta|^2},\quad\quad\quad \epsilon_A = \frac{|t_A|^2}{|r_A|^2}, \quad\quad\quad \epsilon_B \equiv \frac{|t_B|^2}{|r_B|^2}.
\end{align}
From Eq. \ref{eq: output_state}, it follows that the coincidence probabilities for each combination of output ports are proportional to,

\begin{align}
    &C_{A_1, B_1}(\varphi) \propto \sqrt{\frac{\kappa_B}{\kappa_A}}+\sqrt{\frac{\kappa_A}{\kappa_B}}x+2\sqrt{x}\cos{2\varphi},\\
    &C_{A_1, B_2}(\varphi) \propto \frac{1}{\epsilon_B}\sqrt{\frac{\kappa_B}{\kappa_A}}+\epsilon_B\sqrt{\frac{\kappa_A}{\kappa_B}}x+2\sqrt{x}\cos{2\varphi},\\
    &C_{A_2, B_1}(\varphi) \propto \frac{1}{\epsilon_A}\sqrt{\frac{\kappa_B}{\kappa_A}}+\epsilon_A\sqrt{\frac{\kappa_A}{\kappa_B}}x+2\sqrt{x}\cos{2\varphi},\\
    &C_{A_2, B_2}(\varphi) \propto  \frac{1}{\epsilon_A\epsilon_B}\sqrt{\frac{\kappa_B}{\kappa_A}}+\epsilon_A\epsilon_B\sqrt{\frac{\kappa_A}{\kappa_B}}x+2\sqrt{x}\cos{2\varphi},
\end{align}

where the phase factors in the reflectivities $r_A, r_B$ are absorbed into the definition of $\varphi$.  Therefore, the entanglement visibilities, $V = \frac{\text{max}(C(\varphi))-\text{min}(C(\varphi))}{\text{max}(C(\varphi))+\text{min}(C(\varphi))}$,  for each combination of output ports are:
\begin{align}
    &V_{A_1, B_1} = \frac{2\sqrt{x}}{\sqrt{\frac{\kappa_B}{\kappa_A}}+\sqrt{\frac{\kappa_A}{\kappa_B}}x}, \label{eq:FA1B1}\\
    &V_{A_1, B_2} = \frac{2\sqrt{x}}{\frac{1}{\epsilon_B}\sqrt{\frac{\kappa_B}{\kappa_A}}+\epsilon_B\sqrt{\frac{\kappa_A}{\kappa_B}}x},\label{eq:FA1B2}\\
    &V_{A_2, B_1} =  \frac{2\sqrt{x}}{ \frac{1}{\epsilon_A}\sqrt{\frac{\kappa_B}{\kappa_A}}+\epsilon_A\sqrt{\frac{\kappa_A}{\kappa_B}}x},\label{eq:FA2B1}\\
    &V_{A_2, B_2} = \frac{2\sqrt{x}}{\frac{1}{\epsilon_A\epsilon_B}\sqrt{\frac{\kappa_B}{\kappa_A}}+\epsilon_A\epsilon_B\sqrt{\frac{\kappa_A}{\kappa_B}}x}.\label{eq:FA2B2}
\end{align}

\begin{figure}[H]
    \includegraphics[width = \textwidth]{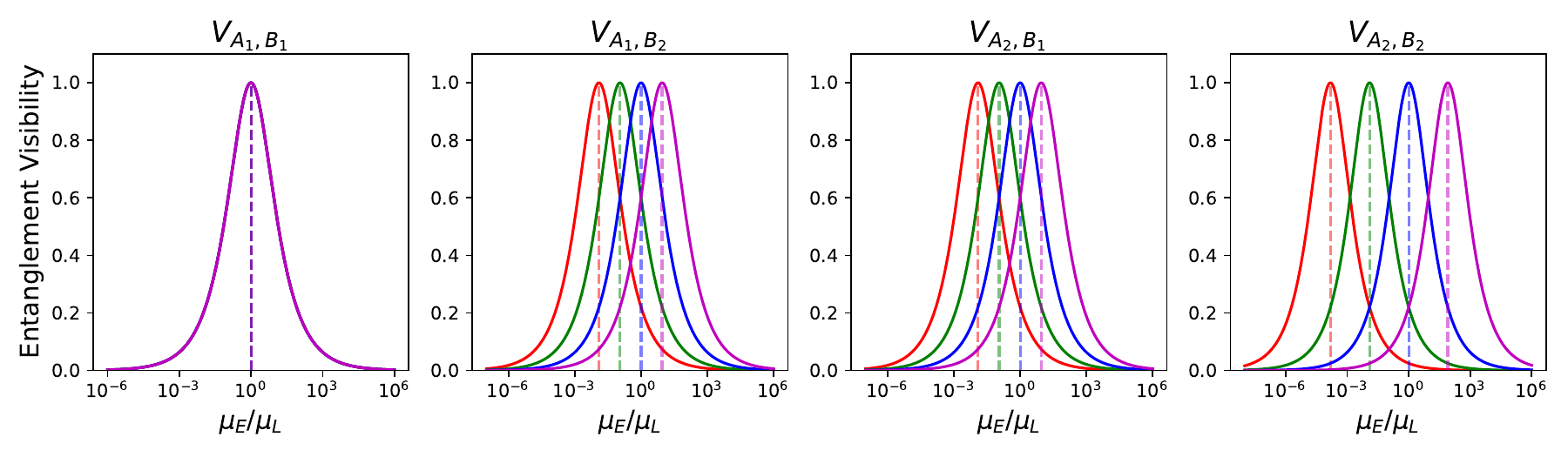}
    \caption{Entanglement visibility as function of $\mu_E/\mu_L$ for fixed $\kappa_B/\kappa_A = 1$ and  $ \epsilon_A = \epsilon_B = 90/10$ (red), $75/25$ (blue), $50/50$ (green), $25/75$ (purple).}
    \label{fig:ent_fid_single_photon}
\end{figure}

 Unity visibility is achievable for each combination of output ports: $V_{A_1, B_1} = 1$ when $x = \kappa_B/\kappa_A$, $V_{A_1, B_2} = 1$ when $x =  \kappa_B/(\kappa_A\epsilon_B^2)$, $V_{A_2, B_1} = 1$ when $x =  \kappa_B/(\kappa_A\epsilon_A^2)$, and $V_{A_2, B_2} = 1$ when $x =  \kappa_B/(\kappa_A\epsilon_A^2\epsilon_B^2)$. Therefore, the effect of imbalances in the source and measurement interferometers is to shift the optimal ratio of early to late mean photon numbers. Imbalances in the measurement interferometers can be compensated by imbalances in the source interferometer in order to obtain unity visibility. Moreover, in the single photon limit, the visibility is insensitive to the absolute path efficiencies in the experiment. The visibility depends only on the ratio of path efficiencies between the measurement interferometers ($\kappa_A/\kappa_B)$. The entanglement visibilities for each combination of output ports as a function of $x=\mu_E/\mu_L$ for various ratios of interferometric path efficiencies are shown in Fig. \ref{fig:ent_fid_single_photon}. 
 
\subsection{Multiphoton Effects}
\begin{figure}[h!]
    \centering
    \includegraphics[width = \textwidth]{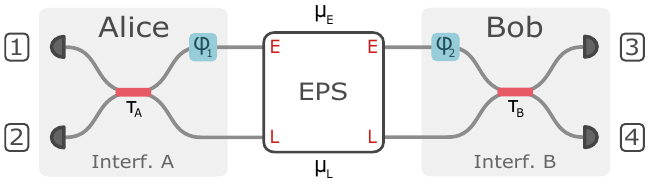}
    \caption{Setup for phase space modeling of entanglement visibility experiment}
    \label{fig:model_setup}
\end{figure}
Calculating the entanglement visibility to higher order photon number contributions quickly becomes intractable with the Fock space approach in section B. To study the effect of multiphoton events on the entanglement visibility, we model the experiment using phase space methods based on a characteristic function formalism \cite{Davis2022}\cite{takeoka2015full}. The model setup is shown in Fig \ref{fig:model_setup}. As in the Fock space approach, the input state is modeled as a product state of TMSV in early and late temporal modes, with mean photon numbers $\mu_E$ and $\mu_L$, respectively. The measurement interferometers are modeled as beamsplitters in the temporal domain that mix the early and late input modes with transmittances $\tau_A$ and $\tau_B$, which absorb the interferometric path efficiencies and spatial beamsplitter transmittances. Since the input state is modeled as a Gaussian state, and the measurement interferometers are modeled as Gaussian operations, we can find the symplectic transformation that maps the characteristic function of the input state to that of the state prior to detection.

\begin{figure}[H]
\centering
    \begin{subfigure}{0.75\textwidth}
        \centering
        \includegraphics[width = \textwidth]{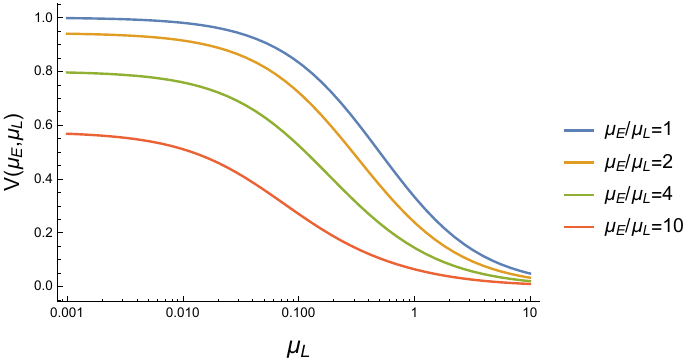}
    \end{subfigure}
    \bigskip
    \begin{subfigure}{0.75\textwidth}
        \centering
        \includegraphics[width = \textwidth]{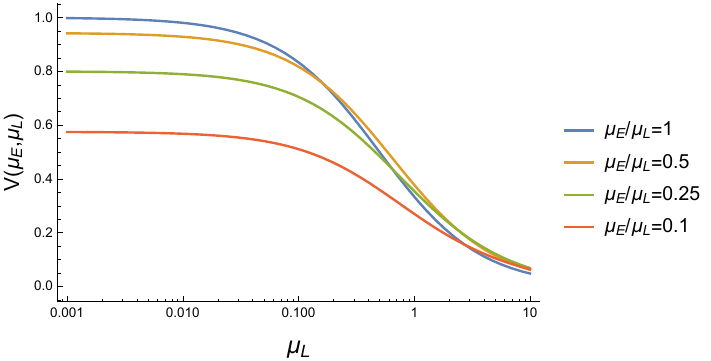}
    \end{subfigure}
    \caption{Entanglement visibility as a function of mean photon number for $\tau_A = \tau_B = 1/\sqrt{2}$.}
    \label{fig:fid_multiphoton}
\end{figure}

Following Ref. \cite{Davis2022}, the characteristic function for an $N$-mode bosonic state is
\begin{align}
    \chi(\xi)=\text{Tr}\left[\hat{\rho}\exp(-i(\hat{x}_1,\hat{p}_1,\hat{x}_2,\hat{p}_2,\cdots\hat{x}_N,\hat{p}_N) \xi)\right]
\end{align}
where $\xi\in \mathbb{R}^{2N}$, $\rho$ is the density matrix, and $\hat{x}_i = \frac{1}{\sqrt{2}}(\hat{a}_i^\dagger+\hat{a}_i)$ and $\hat{p}_i=\frac{1}{\sqrt{2}}(\hat{a}_i^\dagger+\hat{a}_i)$ are the quadrature operators for mode $i$ with annihilation operator $\hat{a}_i$. 
A Gaussian state is a state whose characteristic function that takes a Gaussian form,
\begin{align}
    \chi(\xi) = \exp(-\frac{1}{4}\xi^T\gamma \xi - id^T \xi),
\end{align}
which is fully characterized by the displacement vector $d$ and covariance matrix $\gamma$, i.e. the first and second moments.
 For the TMSV state, the displacement vector is the null vector $d =(0,0,0,0)$ and the covariance matrix is given by
 \begin{align}
     \gamma_{TMSV}(\mu) = 
     \begin{pmatrix}
     \mathbf{A}&\mathbf{B}\\
     \mathbf{B}&\mathbf{A}
     \end{pmatrix}, \quad
     \mathbf{A}=
     \begin{pmatrix}
         1+2\mu&0\\
         0&1+2\mu
     \end{pmatrix}, \quad
     \mathbf{B}=\begin{pmatrix}
         2\sqrt{\mu(\mu+1)}&0\\
         0&-2\sqrt{\mu(\mu+1)}
     \end{pmatrix}
 \end{align}
 where $\gamma_{TMSV}(\mu)$ is written in block matrix form. Therefore, the covariance matrix for the input state of our experiment is
 \begin{align}
     \gamma_{in}(\mu_E,\mu_L) = \gamma_{TMSV}(\mu_E)\oplus \gamma_{TMSV}(\mu_L)
 \end{align}
The characteristic function of the input state is mapped to the characteristic function of the state prior to detection by a symplectic transformation,
\begin{align}
    \chi_{in}(\xi) = \exp(-\frac{1}{4}\xi^T\gamma_{in} \xi)\mapsto \chi_{out}(\xi) = \exp(-\frac{1}{4}\xi^T S^T \gamma_{in}S \xi)
\end{align}
where $S$ is the Symplectic matrix of the interferometers. We construct $S$ from the Symplectic matrices of the phase shifter ($S_{PS}$) and beamsplitter ($S_{BS}$)\cite{takeoka2015full},
\begin{align}
    &S_{PS}(\varphi) = \begin{pmatrix}
        \cos{\varphi}&\sin{\varphi}\\
        -\sin{\varphi}&\cos{\varphi}
    \end{pmatrix},\\
    &S_{BS}(\tau) = 
    \begin{pmatrix}
    \mathbf{T}&\mathbf{R}\\
    \mathbf{R}&\mathbf{T}
    \end{pmatrix}, \quad
    \mathbf{T}=\begin{pmatrix}
        \tau&0\\
        0&\tau
    \end{pmatrix}, \quad \mathbf{R}=
    \begin{pmatrix}
    0&-\sqrt{1-\tau^2}\\
    \sqrt{1-\tau^2}&0
    \end{pmatrix}.
\end{align}
From the output characteristic function $\chi_{\text{out}}$, we obtain the coincidence probabilities using Eq. (9) of Ref. \cite{Davis2022},
\begin{align}
    \text{Tr}\left[\hat{\rho}_{\text{out}}\hat{\Pi}\right] = \left(\frac{1}{2\pi}\right)^N \int {dx}^{2N}\chi_{\text{out}} (x) \chi_\Pi (-x)
\end{align}
where $\hat{\rho}_{\text{out}}$ is the state prior to detection with characteristic function $\chi_{\text{out}}$, and $\hat{\Pi}$ is the measurement operator corresponding to coincidences between detectors from different interferometers. The measurement operators for a threshold detector, which distinguishes between a detection event (at least one photon) and no detection event (zero photons), are
\begin{align}
    \hat{\Pi}_{\text{event}} = \hat{I}-\ket{0}\bra{0}, \quad  \hat{\Pi}_{\text{no event}} = \ket{0}\bra{0}
\end{align}
where $\hat{I}$ is the 2 by 2 identity matrix.
The measurement operator for coincidences between e.g. detectors 1 and 4 are 
\begin{align}
    \hat{\Pi}_{1,4}=\hat{\Pi}_{\text{event},1}\otimes\hat{I}_2 \otimes \hat{I}_3 \otimes \hat{\Pi}_{\text{event},4}
\end{align}
where the subscripts denote the output modes labeled in Fig. \ref{fig:model_setup}. We derive an analytical expression for the coincidence probability that encompasses all multiphoton contributions,
\begin{align}
    C(\varphi) =  \text{Tr}\left[\hat{\rho}_{\text{out}}\hat{\Pi}_{1,4}\right] &= 1-\frac{1}{|f(\mu_E, \mu_L, \tau_A)|}-\frac{1}{|g(\mu_E, \mu_L, \tau_B)|}+\frac{1}{|h(\mu_E, \mu_L, \tau_A, \tau_B,\varphi)|},\\
    f(\mu_E, \mu_L, \tau_A) &= 1+\mu_L+\tau_A(\mu_E - \mu_L),\\
    g(\mu_E, \mu_L, \tau_B) &= 1+\mu_E+\tau_B(\mu_L - \mu_E),\\
    h(\mu_E, \mu_L, \tau_A, \tau_B) &=1+\mu_E + \mu_L(1 + \mu_E)(1 -\tau_A) \\
    &-\mu_E\tau_B(1 + \mu_L) +\tau_A\tau_B(\mu_E  + \mu_L +2\mu_E\mu_L) \nonumber\\
    &- 2\sqrt{\mu_E\mu_L\tau_A(1+\mu_E)(1+\mu_L)(1-\tau_A)}\sqrt{\tau_B(1-\tau_B)}\cos{\varphi},\nonumber
adfs\end{align}
where $\varphi = \varphi_A-\varphi_B$ is the relative phase between interferometers A and B.
 The different visibilities in each output port combination as a result of interferometric imbalances can be obtained by adjusting $\tau$ accordingly. To isolate the impact of multiphoton contributions to the visibility, we set $\tau_A = \tau_B = \frac{1}{\sqrt{2}}$, and obtain the following expression for the entanglement visibility:


 \begin{align}
    V(\mu_E, \mu_L) &= \frac{C(0)-C(\pi)}{C(0)+C(\pi)}=\frac{2/\sqrt{G_{-}(\mu_E, \mu_L)}-2/\sqrt{G_{+}(\mu_E, \mu_L)}}{1-4/(2+\mu_E+\mu_L)+2/\sqrt{G_{-}(\mu_E, \mu_L)}+2/\sqrt{G_{+}(\mu_E, \mu_L)}}\label{eq:fid_multiphoton}
\end{align}

\begin{align}
     G_{\pm}(\mu_E, \mu_L)&=\mu_E^2(9+8\mu_L(2+\mu_L))\\
     &\pm(4+3\mu_L)\bigg(\pm4\pm3\mu_L+4\sqrt{\mu_E \mu_L (1+\mu_E)(1+\mu_L)}\bigg)\nonumber\\     &+2\mu_E\bigg(12\pm6\sqrt{\mu_E\mu_L(1+\mu_E)(1+\mu_L)}\bigg)\nonumber\\
     &+2\mu_E\mu_L\Big(19+8\mu_L \pm 4\sqrt{\mu_E\mu_L (1+\mu_E)(1+\mu_L)}\Big). \nonumber
 \end{align}
 By expanding Eq. \ref{eq:fid_multiphoton} to first order in $\mu_E$ and $\mu_L$,
 \begin{align}
     V(\mu_E,\mu_L) = \frac{2\sqrt{\frac{\mu_E}{\mu_L}}}{1+\frac{\mu_E}{\mu_L}} - \frac{\mu_E}{\mu_L}\frac{\left(5(\frac{\mu_E}{\mu_L}+\frac{\mu_L}{\mu_E})+6\right)}{2(1+\frac{\mu_E}{\mu_L})^2}\sqrt{\mu_E\mu_L}+\cdots\label{eq: fid_first_order}
 \end{align}
 we see that the first term matches Eq. \ref{eq:FA1B1}-\ref{eq:FA2B2} for $t_A = t_B = \frac{1}{\sqrt{2}}$, $\beta/\alpha = \gamma/\delta = 1$.  Moreover, for $\mu_{eq} \equiv \mu_E = \mu_L$, Eq. \ref{eq: fid_first_order} reduces to $ V(\mu_{eq}) = 1 - 2\mu_{eq} $. 
Thus, the upper bound on the visibility is set by the mean photon number, i.e. multiphoton effects. Entanglement visibilities of more than 90\% are possible when $0.39<\mu_E/\mu_L <2.55$ and $\mu_L < 0.0.056$.  The entanglement visibility $V(\mu_E, \mu_L)$ in Eq. \ref{eq:fid_multiphoton} is plotted for various mean photon numbers in Fig. \ref{fig:fid_multiphoton}.

\section{Maximum Entangled Photon Source Throughput}

\begin{figure}[H]
    \centering
    \includegraphics[width=1\linewidth]{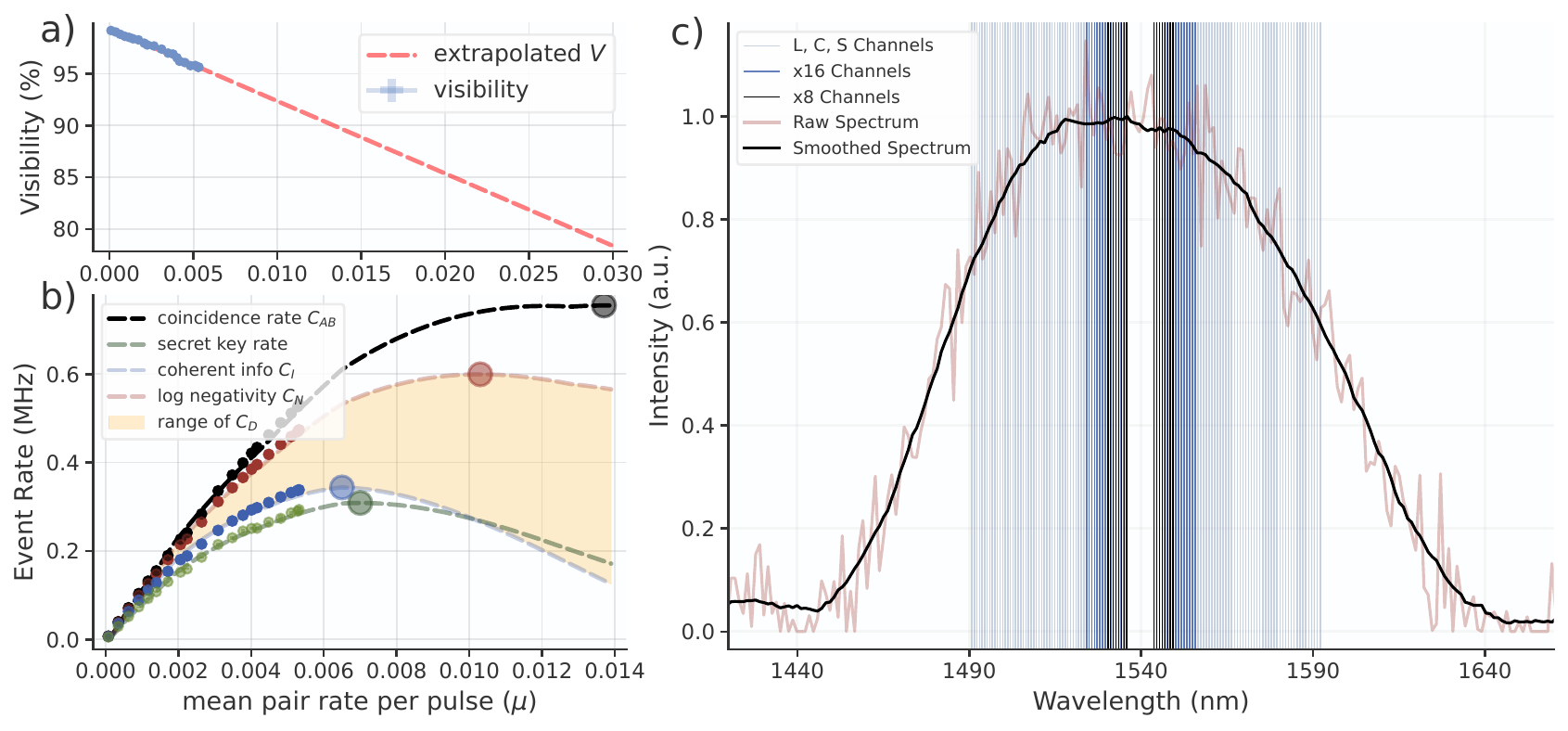}
    \caption{a) and b) include data from main text Figure 3, and extrapolate it to higher $\mu$. It is notable that $C_I$ and $SRK$ reach their maximum values for only marginally higher pump powers than those used for our measurements. c) Spectrum of entangled photon pairs out of the SPDC, measured with an Anritsu MS9740B spectrum analyzer in High Dynamic Range mode. As the signal was near the instrument noise floor, many repeated spectrum measurements were point-averaged.}
    \label{fig:scan_extrapolate}
\end{figure}

We observe in the small $\mu$ limit that the metrics $V$, $E_N = C_N/C_{AB}$, $E_I = C_I/C_{AB}$, and $E_S = SKR/C_{AB}$ scale linearly with $\mu$, where $E_S$ is the secret key fraction. Raw coincidence rate is not linear with $\mu$ due to the count rate dependent efficiency of the SNSPDs. As count rate increases, the detector spends a larger fraction of time in a partially reset state where photo-detection is less efficient or not possible. We separately collect measurements of detector efficiency versus count rate extending past 10~Mcps, and use this to extrapolate coincidence rate to higher powers. Then, the metrics $E_N$, $E_I$, and $E_S$ are multiplied by the extrapolated rate to define extrapolated $C_N$, $C_I$, and $SKR$ as shown in \ref{fig:scan_extrapolate}b. Maximum values of these metrics are highlighted by colored circular markers.

In this work we primarily study the capability of 8 DWDM channel pairs, with some analysis of 16-pair performance. However, the spectrum of entangled photon pairs produced by the type-0 SPDC is quite broad, meaning spectral multiplexing across many more channels is possible. In \ref{fig:scan_extrapolate} the spectrum of the SPDC is overlayed with sixty 100~GHz DWDM channel pairs spanning the L, C, and S ITU bands [L-29 (1592.1~nm) through C-20 (1543.73) and C-10 (1535.82) through S-1 (1490.76)]. The coincidence rate through these added channels can be estimated from this spectrum and the known performance of the central 8 channel pairs. Table 1 in the main text specifies estimates for the total rate from the 60 multiplexed spectral channels.

\section{Heralded $\boldsymbol{g^2(0)}$}

\begin{figure}[H]
    \centering
    \includegraphics[width=1\linewidth]{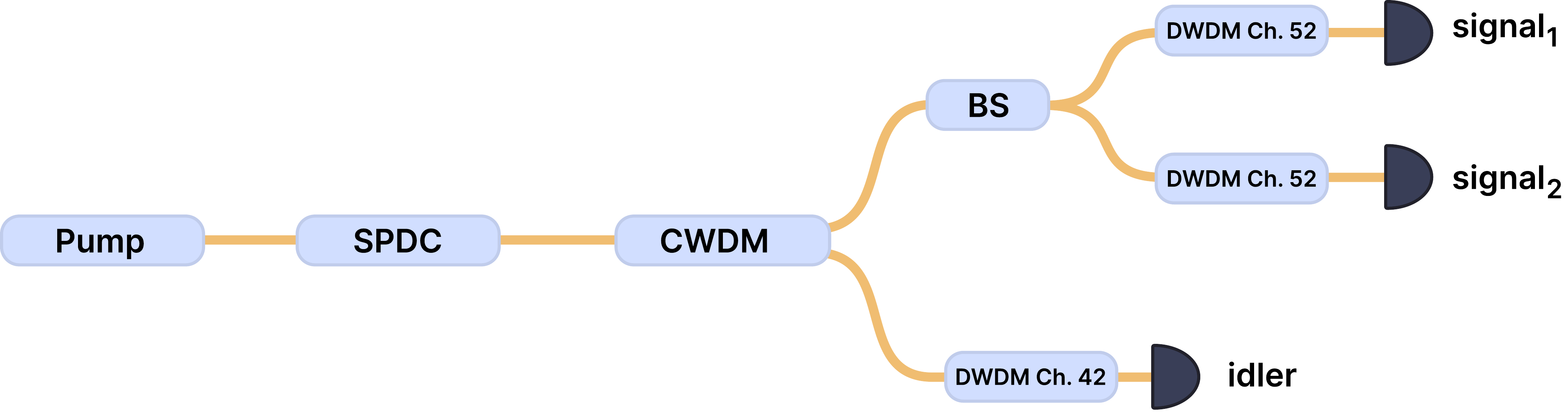}
    \caption{Heralded $g^2(0)$ configuration}
    \label{fig:heralded_g2}
\end{figure}

To express the heralded $g^2(0)$ of this source, we consider an alternate measurement configuration in \ref{fig:heralded_g2}. Here, we imagine using just one pair of DWDM channels, but the signal arm is resolved with two detectors after a beamsplitter. This is analogous to a configuration in Davis et al. \cite{Davis2022} for which $g^2(0)$ of a pair generating source was studied. We use expressions (3) - (7) from this paper together to make an expression for $g^2(0)$ valid in the small $\mu$ limit. Variables from Davis et al. are shown in blue. Davis et al. uses $\textcolor{blue}{C_{i(s)}}$ for idler (signal) singles rate, while here we use $S_{i(s)}$. Note that $\textcolor{blue}{\mu}$ is different from $\mu$ as defined in the main text, since Davis et al. paper does not operate in a narroband filtering regime. 

With equations (3) - (7) from Davis et al.,

\begin{align}
    &\textcolor{blue}{(3)\,\,\,\,C_i \approx R \eta_i \mu } \label{eq:start}\\ 
    &\textcolor{blue}{(4)\,\,\,\,C_{s_j} \approx \frac{1}{2} R \eta_{s_j} \mu} \\
    &\textcolor{blue}{(5)\,\,\,\,C_{i s_j} \approx \frac{1}{2} R \eta_i \eta_{s_j} \mu} \\
    &\textcolor{blue}{(6)\,\,\,\,C_{s_1 s_2} \approx \frac{1}{2} R \eta_{s_1} \eta_{s_2} \mu^2 } \\
    &\textcolor{blue}{(7)\,\,\,\,C_{i s_1 s_2} \approx \frac{1}{2} R \mu^2 \eta_i \eta_{s_1} \eta_{s_2}\left(2-\eta_i\right)} \label{eq:end}
\end{align}


$g^2(0)$ for the configuration in \ref{fig:heralded_g2} may be written as

$$ \textcolor{blue}{g^2(0)=\frac{C_{i s_1 s_2} C_i}{C_{i s_1} C_{i s_2}}.} $$

Substituting in the expressions on the right-hand side of \ref{eq:start} - \ref{eq:end}, the expression for $g^2(0)$ becomes

$$ \textcolor{blue}{g^2(0)=2(2 - \eta_i)\mu} \rightarrow \frac{2 S_i(2 - \eta_i)}{R \eta_i} $$

\begin{figure}[H]
    \centering
    \includegraphics[width=0.8\linewidth]{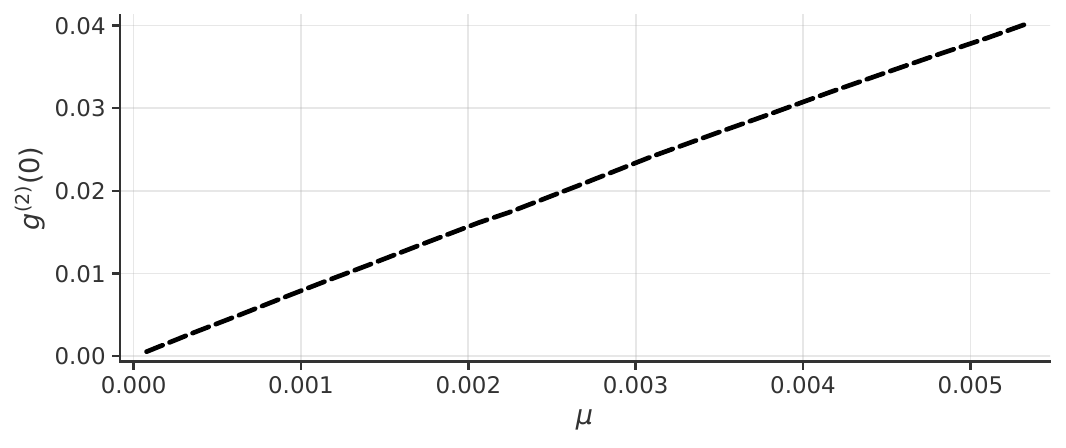}
    \caption{Given the small $\mu$ limit, $g^2(0)$ is approximately linearly dependent on $\mu$}
    \label{fig:g2}
\end{figure}

Where the last transformation assumes $\textcolor{blue}{\mu} = \frac{S_i}{R \eta_i}$. We may insert data from this source and graph $g^2(0)$ vs $\mu$ as shown in \ref{fig:g2}. We use $\eta_i = 0.17$ from Alice's channel 42 in the main text figure 2a. Given our unique defenition of $\mu$, we find $g^2(0) \approx 7.7 \mu$ in the small $\mu$ limit.

\bibliography{references}